\documentclass[12pt]{amsart}
\usepackage{amsmath,amsfonts,latexsym,graphicx,amssymb,url}
\usepackage{amsmath,txfonts,pifont,bbding,pxfonts,manfnt}
\usepackage{psfrag}
\usepackage{empheq}
\usepackage{color}
\usepackage{mathrsfs}
\usepackage{wrapfig, subfigure,graphicx}
\usepackage{hyperref}
\usepackage[active]{srcltx}
\usepackage{breqn}
\usepackage{pdfsync}
\usepackage{pdflscape}
\newcommand{\R}{{\mathbb R}} %%reals

\theoremstyle{plain}
\newtheorem{theorem}{Theorem}[section]

\newtheorem{remark}[theorem]{Remark}

\begin{document}

\setcounter{equation}{0}

%\documentstyle[12pt]{article}

%\usepackage{amsmath}
%\usepackage{amsfonts}
%\usepackage{amssymb}
%\usepackage{amscd}

%\setlength{\headheight}{15pt}
%\setlength{\topmargin}{10pt}
%\setlength{\headsep}{30pt}
%\setlength{\textwidth}{15cm}
%\setlength{\textheight}{21.5cm}
%\setlength{\oddsidemargin}{1cm}
%\setlength{evensidemargin}{1cm}

%\newtheorem{theorem}{Theorem}[section]

%\newtheorem{lemma}[theorem]{Lemma}

%\newtheorem{corollary}[theorem]{Corollary}

%\newtheorem{proposition}[theorem]{Proposition}

%\newtheorem{definition}[theorem]{Definition}

%\begin{document}

\title[General Refraction Problems with Phase Discontinuity]
{General Refraction Problems with Phase Discontinuity}
\author[C. E. Guti\'errez, L. Pallucchini, and E. Stachura]
{Cristian E. Guti\'errez, Luca Pallucchini, and Eric Stachura}
\thanks{\today\\The first author is partially supported
by NSF grant DMS--1600578.}
\address{Address for C.E.G and L.P: Department of Mathematics\\Temple University\\Philadelphia, PA 19122}
\email{gutierre@temple.edu, luca.pallucchini@temple.edu}
\address{Department of Mathematics\\Haverford College\\Haverford, PA 19041}
\email{estachura@haverford.edu}

\maketitle
\begin{abstract}
This paper provides a mathematical approach to study metasurfaces in non flat geometries. Analytical conditions between the curvature of the surface and the set of refracted directions are introduced to guarantee the existence of phase discontinuities. The approach contains both the near and far field cases.
A starting point is the formulation of a vector Snell law in presence of abrupt discontinuities on the interfaces.
\end{abstract}
\tableofcontents

\section{Introduction}
For classical lens design, a typical problem is to find two surfaces so that the region sandwiched between them and filled with an homogeneous material refracts light in a desired manner. 
For metalens design, a surface is given and the question is to find a function on the surface (a phase discontinuity) so that the pair, surface together with the phase discontinuity (the metalens) refracts light in a desired manner. 
The subject of metalenses is a flourishing area of research and one of the nine runners-up for Science's Breakthrough of the Year 2016 \cite{science-runner-ups-2016}. Metalenses have been designed for flat geometries with the scalar generalized laws of reflection and refraction
with phase discontinuities, see \cite{yu2011light}, \cite{aieta2012out},  \cite{aieta2012reflection}, and the comprehensive review article \cite{yu-capasso:flatoptics}.
 These general laws have been experimentally observed by using arrays of optical antennas on silicon. The review in \cite{chen2016review} describes the past 15 years of progress on metasurfaces, from experimental realization of the generalized laws of refraction, to applications in wavefront and beam shaping. Recently, it has been proven \cite{ammari2016mathematical} that at certain frequencies, a thin layer of nanoparticles on a perfectly conducting sheet acts as a metasurface.
For more recent work in the area and an extensive up to date bibliography, we refer to \cite{2107planaroptics:capasso}; see also  \cite{2016metalensesvisiblewavelenghts:capasso} and \cite{2016visiblewavelenghtsdiffraction:capasso}.
  
 %To our knowledge this is one of the first rigorous mathematical works in this area. 

\begin{figure}
\includegraphics[width=3in]{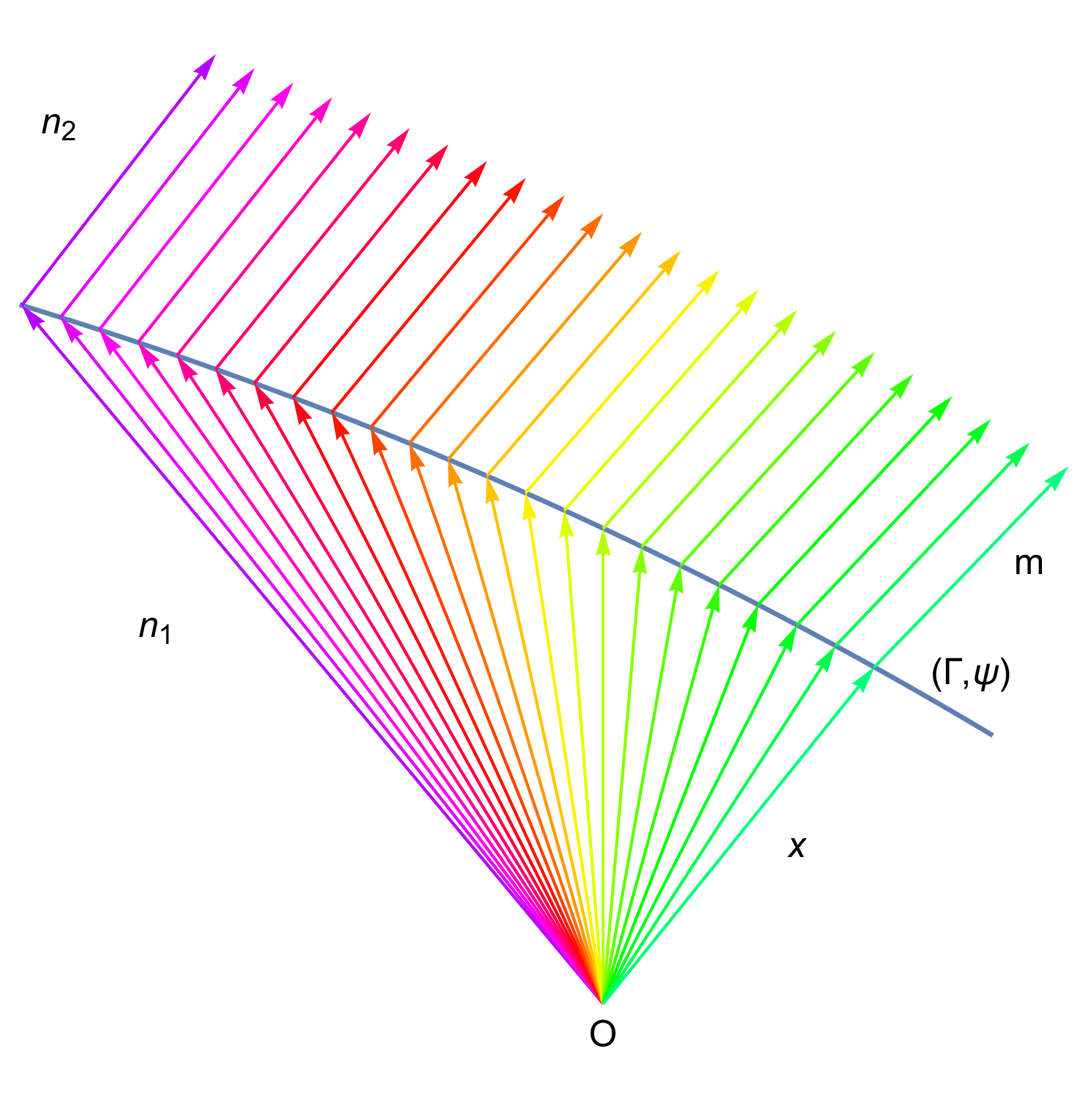}
\caption{Metalens refracting into a fixed direction}
\label{fig:general problem}
\end{figure}

%For classical lens design, a typical problem is to find two surfaces so that the region sandwiched between them and filled with an homogeneous material refracts light in a desired manner. 
%For metalens design, a surface is given and the question is to find a function on the surface, a phase discontinuity, so that the pair, surface together with the phase discontinuity, the metalens, refracts light in a desired manner. 
%The subject of metalenses is one of the nine runners-up for Science's Breakthrough of the Year 2016 \cite{science-runner-ups-2016}. Metalenses have been designed for simple geometries based on the scalar generalized laws of reflection and refraction
%with phase discontinuities, see \cite{yu2011light}, \cite{aieta2012out}, and \cite{aieta2012reflection}. 
The purpose of this paper is to provide a mathematically rigorous foundation to deal with general metasurfaces and to determine the relationships between the curvature of the surface and the phase discontinuity. 
A problem we solve is the following: when light emanates from a point source, find a metalens that refracts light into a prescribed set of directions or points, see Figure \ref{fig:general problem}.
In fact, given a surface and a compatible set of directions, satisfying appropriate curvature type conditions, we show that a phase discontinuity exists on the surface so that the metalens refracts light into the prescribed set of directions, Section \ref{sec:m variable}. Vice versa, given a phase discontinuity and a fixed direction, we find the admissible surfaces for that phase discontinuity and direction, Section \ref{subsec:given the field find the phase discontinuity}. Of great importance to answer these questions in general geometries is the formulation of a generalized Snell's law in vector form, Equation \eqref{vect2}, which is deduced using wave fronts in Section \ref{sec:derivationofsnelllawwithfasediscontinuities}.
In term of wave vectors, a vector law is formulated in \cite[Equation (2)]{aieta2012reflection}. However, Equation \eqref{vect2} is effective and flexible for the actual
calculation of phase discontinuities in general and to obtain our results. 
We illustrate these with explicit constructions for planar and spherical interfaces, Sections \ref{subsec:case of the plane}, \ref{subsec:caseofthespherecenteredatzero}, \ref{subsec:near field plane case}, and \ref{subsec:near field spherical case}; see also Remark \ref{rmk:aieta case is incorrect}.

%This was accomplished by using Fermat's principle as a principle of stationary phase. The results are formulated in terms of incident and refracted wave vectors; motivated by optical interface problems where a vector form of Snell's law is required in terms of the refractive indices of the media (see e.g. \cite{Gutinotes}), we formulate an equivalent version of Snell's law in the presence of phase discontinuities which depends on the refractive indices of the materials involved. 
%
%

The outline of the paper is as follows. In Section \ref{sec:classical snell law}, we briefly recall the classical Snell's law for surfaces without phase discontinuities. Then in Section \ref{sec:derivationofsnelllawwithfasediscontinuities} we derive a generalized Snell's law in the presence of a phase discontinuity using wavefronts, Equation (\ref{vect2}), and analyze the possible critical angles. The far field problem is studied in Section \ref{sec:Far field uniformly refracting surfaces} for the plane and the sphere. In Section \ref{sec:m variable} we allow for variable directions $m$ in the far field. In Section \ref{subsec:given the field find the phase discontinuity}, conditions are derived so that given a phase discontinuity a surface exists. Finally, in Section \ref{sec:near field case} the near field problem is addressed.

\section{Background}\label{sec:classical snell law}
\setcounter{equation}{0}
We recall the classical Snell's law in vector form here. Suppose $\Gamma$ is a surface in $\mathbb{R}^3$ that separates two media
$I$ and $II$ that are homogeneous and isotropic, with refractive indices $n_1$ and $n_2$ respectively.
%Let $v_1$ and
%$v_2$ be the velocities of propagation of light in the media I and
%II respectively. The index of refraction of the medium I is by
%definition $n_1=c/v_1$, where $c$ is the velocity of propagation
%of light in vacuum, and similarly $n_2=c/v_2$. 
If a
ray of light\footnote{Since the refraction angle depends on the frequency of the radiation, we assume that light rays are monochromatic.} having direction $x\in S^{2}$, the unit sphere in $\mathbb{R}^3$, and traveling
through medium $I$ strikes $\Gamma$ at the point $P$, then this ray
is refracted in the direction $m\in S^{2}$ through medium $II$
according to the Snell law in vector form:
\begin{equation}\label{snellwithcrossproduct}
n_{1}(x\times \nu)=n_{2}(m\times \nu),
\end{equation} 
where $\nu$ is the unit normal to the surface to $\Gamma$ at $P$ pointing towards medium $II$; see \cite[Subsection 4.1]{luneburgoptics}. It is assumed here that $x\cdot \nu\geq 0$.

This has several consequences:
\begin{enumerate}
\item[(a)] the vectors $x,m,\nu$ are all on the same plane (called the plane of incidence);
\item[(b)] the well known Snell's law in scalar form holds: 
$$n_1\sin \theta_1= n_2\sin
\theta_2,$$ 
where $\theta_1$ is the angle between $x$ and $\nu$
(the angle of incidence), and
$\theta_2$ is the angle between $m$ and $\nu$ (the angle of refraction).
\end{enumerate}
Equation \eqref{snellwithcrossproduct} is equivalent to $(n_{1}x-n_{2}m)\times \nu=0$, which means that the
vector $n_{1}x-n_{2}m$ is parallel to the normal vector $\nu$.
If we set $\kappa=n_2/n_1$, then
\begin{equation}\label{eq:snellvectorform}
x-\kappa \,m =\lambda \nu,
\end{equation}
for some $\lambda\in \mathbb{R}$. Notice that \eqref{eq:snellvectorform} univocally determines $\lambda$. Taking dot products with $x$ and $m$ in \eqref{eq:snellvectorform} we get
$\lambda=\cos \theta_1-\kappa \cos \theta_2$,
$\cos \theta_1=x\cdot \nu>0$, and
$\cos \theta_2=m\cdot \nu=\sqrt{1-\kappa^{-2}[1-(x\cdot \nu)^2]}$. In fact, there holds
\begin{equation}\label{eq:formulaforlambda}
\lambda=x\cdot \nu -\kappa \,\sqrt{1-\kappa^{-2}\left(1-(x\cdot \nu)^{2}\right)}.
\end{equation}

The formulation \eqref{eq:snellvectorform} is useful solve refraction problems for lens design, see 
\cite{gutierrez-huang:farfieldrefractor},
\cite{gutierrez-mawi:refractorwithlossofenergy},
\cite{gutierrez-sabra:thereflectorproblemandtheinversesquarelaw},
\cite{gutierrez-tournier:parallelrefractor}, and \cite{deleo-gutierrez-mawi:numericalalgorithm} for a numerical implementation.

\section{Derivation of a Vector Snell Law with phase discontinuity using wavefronts}\label{sec:derivationofsnelllawwithfasediscontinuities}
\setcounter{equation}{0}
Let $n_1, n_2$ be the refractive indices of two homogeneous media $I$ and $II$, respectively. Suppose a surface $\Gamma$ separates the two media, and an incoming light ray in medium $I$ with wave vector $\textbf{k}_1$ strikes $\Gamma$. 
Assume that there is a real-valued function $\psi$, {\it the phase discontinuity}, defined in a neighborhood of the surface $\Gamma$.
Notice that $\psi$ must be defined in a neighborhood of $\Gamma$ because the gradient of 
$\psi$ will be considered.
%WHAT IS THE MEANING OF THIS SENTENCE?:Suppose that $\Gamma$ has a constant phase gradient $\nabla \psi$ along an arbitrary direction on the surface. 
If $\nu$ denotes the unit normal vector to $\Gamma$, then the refracted wave vector $\textbf{k}_2$ satisfies \cite[Equation (2)]{aieta2012reflection}:
%The direction of the refracted ray, having wave vector $\textbf{k}_2$, can be found via the Stationary Phase principle as discussed above.
\begin{align} \label{vectorSnellwithwavevector}
\nu \times (\textbf{k}_2 - \textbf{k}_1) = \nu \times \nabla \psi
\end{align}
We give an alternate formulation and derivation of this result by using wavefronts; our starting point is \cite[Section 2.2]{Gutinotes}. For each $t$, $\Psi(x,y,z,t)=0$ denotes a surface in the variables $x,y,z$ that separates the part of the space that is at rest from the part of the space that is
disturbed by the electric and magnetic fields. This surface is called a {\it wave front}, and
%For each $t$ fixed the surface defined
%by $\Psi(x, y, z, t) = 0$ is called a wave front. 
the light rays are the orthogonal trajectories
to the wave fronts at each time $t$. We assume that $\Psi_t \neq 0$, and so we can solve $\Psi(x, y, z, t) = 0$ in $t$, obtaining that $\phi(x,y,z) = ct$; so letting $t$ run, the wave fronts %\footnote{In terms of the wave vector $\textbf{k}$, the wavefront is the surface defined by $\textbf{k} \cdot \textbf{r} = \text{constant}$, where $\textbf{r} = (x,y,z)$.} 
are then the level sets of $\phi(x,y,z)$. 

Let $n_1, n_2$, and $\Gamma$ be as above. An incoming wave front $\Psi_1$ on medium $I$ strikes the surface $\Gamma$ and it is then transmitted into a wave front $\Psi_2$ in medium $II$ (of course, there is also a wave front reflected back). Assuming as before that $(\Psi_j)_t\neq 0$, $j=1,2$, and solving in $t$, we get that the wave fronts are given by $\phi_j(x,y,z) = ct$ for $j=1,2$, respectively. Suppose the surface $\Gamma$ is parameterized by $x= f(\xi, \eta)$, $y = g(\xi, \eta)$, $z = h(\xi, \eta)$. If there were no phase discontinuity on the surface $\Gamma$, then we would have $\phi_1 = \phi_2$ along $\Gamma$. But since there is now a phase discontinuity $\psi$ on $\Gamma$, we have the following jump condition along $\Gamma$:
$$\phi_1(f(\xi, \eta), g(\xi, \eta), h(\xi, \eta)) - \phi_2(f(\xi, \eta), g(\xi, \eta), h(\xi, \eta)) = \psi(f(\xi, \eta), g(\xi, \eta), h(\xi, \eta)).$$
%HERE $\psi$ IS A GIVEN FUNCTION DIFFERENT FROM THE $\psi(x,y,x,t)$ INTRODUCED BEFORE.
Taking derivatives in $\xi$ and $\eta$ yields
$$\left( \nabla \phi_1 - \nabla \phi_2 - \nabla \psi \right) \cdot (f_\xi, g_\xi, h_\xi) =0,$$ 
and
$$\left( \nabla \phi_1 - \nabla \phi_2 - \nabla \psi \right) \cdot (f_\eta, g_\eta, h_\eta) =0.
$$
That is, the vector $\nabla \phi_1 - \nabla \phi_2 - \nabla \psi$ must be normal to $\Gamma$; as such there exists a real number $\lambda$ such that
\begin{align} \label{vect1}
\nabla \phi_1 - \nabla \phi_2 - \nabla \psi = \lambda \nu
\end{align}
where $\nu$ is the unit normal to $\Gamma$.

Let $\gamma_j(t)$ denote the light rays in medium $j$ having speed $v_j$, for $j=1,2$; i.e., the orthogonal trajectories to $\phi_j$. In particular, we have that $\phi_j (\gamma_j(t)) = ct$, and by the chain rule
$$\nabla \phi_j (\gamma_j(t)) \cdot \gamma_j'(t) = c, \quad j=1,2$$
If we parameterize the rays so that $|\gamma_j'(t)| = v_j$, then we obtain
$$|\nabla \phi_j (\gamma_j(t))| = \dfrac{c}{v_j} = n_j, \quad j=1,2$$
since $\nabla \phi_j$ is parallel to $\gamma_j'$.
Letting 
$$x = \dfrac{\nabla \phi_1(\gamma_1(t))}{| \nabla \phi_1(\gamma_1(t))|}, \quad m = \dfrac{\nabla \phi_2(\gamma_2(t))}{| \nabla \phi_2(\gamma_2(t))|}$$ 
we obtain from \eqref{vect1} the following formula 
\begin{align}\label{vect2}
n_1 x -n_2 m = \lambda \nu + \nabla \psi.
\end{align}
Taking cross products with the unit normal $\nu$ in (\ref{vect2}), we obtain the equivalent formula
\begin{align} \label{vect3}
\nu \times (n_1 x - n_2 m) = \nu \times \nabla \psi.
\end{align}
Recall that $x$ is the unit direction of the incident ray, $m$ is the unit direction of the refracted ray, $\nu$ is the unit outer normal at the incident point on $\Gamma$ and $\nabla \psi$ is calculated at the incident point.
Note that in the case $\psi$ is constant, we recover the classical Snell's law in vector form (\ref{snellwithcrossproduct})\footnote{Notice that if $\psi=$constant, then $n_1\,\nu\times x=n_2\,\nu\times m$. Taking dot product with $m$ yields $n_1\,m\cdot (\nu\times x)=0$. This means that $m$ is on the plane through the origin having normal $\nu \times x$ which is the plane generated by $\nu$ and $x$. Therefore $\nu,x,m$ are all on the same plane, i.e., the plane of incidence. On the other hand, if $\psi$ is not necessarily constant, then from \eqref{vect3} $n_1\,\nu\times x=n_2\,\nu\times m+ \nu \times \nabla \psi$. Again taking dot product with $m$ yields $n_1\,m\cdot (\nu\times x)=m\cdot (\nu\times \nabla \psi)$, that is,
$m\cdot \left(\nu \times \left(n_1\,x-\nabla \psi \right) \right)=0$. That is, now the refracted vector $m$ lies on the plane through the origin and perpendicular to the vector $\nu\times \left( n_1\,x-\nabla \psi\right)$ where $\nabla \psi$ is calculated at the point on the surface $\Gamma$ where the ray with direction $x$ strikes it. This shows that in the general case the refracted vector $m$ is not on the plane generated by $\nu$ and $x$.}. 

Starting from \eqref{vect2}, we now calculate $\lambda$.
Taking dot products in \eqref{vect2} and solving for $x\cdot m$ yields
\[
x\cdot m=\dfrac{n_1-\lambda\,x\cdot \nu -x\cdot \nabla \psi}{n_2}.
\]
Next taking dot products in \eqref{vect2} with itself, expanding, and substituting $x\cdot m$ from the previous expression, yields that $\lambda$ satisfies the quadratic equation:
\begin{equation} \label{quadraticforlambda}
 \lambda^2 -\left[2(n_1 x -\nabla \psi ) \cdot \nu\right] \lambda + |n_1 x -\nabla \psi|^2 - n_2^2 = 0.
 \end{equation}
Solving for $\lambda$ yields 
 \begin{dmath} \label{formulaforlambda}
 \lambda = (n_1x - \nabla \psi)\cdot \nu \pm \sqrt{ n_2^2 - \left( |n_1 x -\nabla \psi|^2 - \left[ (n_1 x -\nabla \psi)\cdot \nu \right]^2\right) }.
\end{dmath}
%Notice that
%$$ |n_1 x -\nabla \psi|^2 - \left[ (n_1 x -\nabla \psi)\cdot \nu \right]^2 \geq 0$$
Since $\lambda$ must be a real number, the quantity under the square root must be non-negative, i.e., 
 \begin{equation}\label{eq:conditionondiscriminant}
 n_2^2 \geq |n_1 x -\nabla \psi|^2 - \left[ (n_1 x -\nabla \psi)\cdot \nu \right]^2.
 \end{equation}
 %This condition imposes a growth condition on $\nabla \psi$ because setting $v=n_1 x -\nabla \psi$, then \eqref{eq:conditionondiscriminant} reads 
 %\[
 %\dfrac{n_2^2}{|v|^2}\geq 1-\left(\dfrac{v}{|v|}\cdot \nu \right)^2 
 %\]
 %which is impossible if $|v|$ is sufficiently large, that is, if $\nabla \psi$ is large.
%CAN WE READ DOUBLE CRITICAL ANGLES FROM THIS?
%I THINK WE SHOULD ANALYZE THE POSSIBLE CONFIGURATIONS FOR THIS KIND OF REFRACTION TO HAPPEN.
Assuming this for now, it remains to check which sign ($\pm$) to take in \eqref{formulaforlambda}. 
Dotting (\ref{vect2}) with $\nu$ and using \eqref{formulaforlambda} yields
 \begin{dmath*}
 n_1 x \cdot \nu - n_2 m \cdot \nu =  (n_1x - \nabla \psi)\cdot \nu \pm \sqrt{ n_2^2 - \left( |n_1 x -\nabla \psi|^2 - \left[ (n_1 x -\nabla \psi)\cdot \nu \right]^2\right) }+ \nabla \psi \cdot \nu,
 \end{dmath*}
 so
 \begin{dmath*}
 -n_2 m \cdot \nu = \pm \sqrt{ n_2^2 - \left( |n_1 x -\nabla \psi|^2 - \left[ (n_1 x -\nabla \psi)\cdot \nu \right]^2\right) }.
 \end{dmath*}
Since $n_2>0$ and $m \cdot \nu \geq 0$, we obtain that
\begin{equation}\label{lambdadependsonsign}
 \lambda = 
 (n_1 x - \nabla \psi ) \cdot \nu - \sqrt{ n_2^2 - \left( |n_1 x -\nabla \psi|^2 - \left[ (n_1 x -\nabla \psi)\cdot \nu \right]^2\right) }. 
 \end{equation} 
 
 We next analyze \eqref{eq:conditionondiscriminant}, which will yield the critical angles.
%Suppose first that $\nabla \psi$ is tangential to $\Gamma$, i.e., $\nabla \psi \cdot \nu = 0$.
%Hence \eqref{eq:conditionondiscriminant} is equivalent to 
%\begin{dmath*}
%(x \cdot \nu)^2\geq \left| x-\dfrac{\nabla \psi}{n_1}\right|^2-\kappa^2,
%\end{dmath*}
%where $\kappa =\dfrac{n_2}{n_1}$. 
%If $x$ is such that
%\[
%\left| x-\dfrac{\nabla \psi}{n_1}\right|^2-\kappa^2\leq 0,
%\]
%then \eqref{eq:conditionondiscriminant} holds. 
%On the other hand, if
%\[
%\left| x-\dfrac{\nabla \psi}{n_1}\right|^2-\kappa^2> 0
%\] 
%then \eqref{eq:conditionondiscriminant} holds when 
%\begin{dmath*}
%x \cdot \nu \geq \sqrt{\left| x-\frac{\nabla \psi}{n_1}\right|^2 -\kappa^2}.
%\end{dmath*}
%From this inequality, the critical angles between $x$ and $\nu$ are $\theta_c =\angle (x,\nu)$ with 
%\begin{equation*}
%x \cdot \nu =\cos \theta_c= \sqrt{\left| x-\frac{\nabla \psi}{n_1}\right|^2 -\kappa^2}.
%\end{equation*} 
%%The critical angles therefore exist if and only if $\left| x-\frac{\nabla \psi}{n_1}\right|^2 -\kappa^2 \geq 0$, for
%%$\left| x-\frac{\nabla \psi}{n_1}\right|^2 -\kappa^2 < 0$ we don't have critical angles.
%
%
%Suppose not that $\nabla \psi$ is not necessarily tangential to $\Gamma$. 
Equation \eqref{eq:conditionondiscriminant} is equivalent to 
\[
\left(\left( x  -  \dfrac{\nabla \psi}{n_1}\right)\cdot \nu \right)^2\geq \left|x- \dfrac{\nabla \psi}{n_1}\right|^2-\kappa^2.
\]
Thus, if $x$ is such that
\[
\left| x-\dfrac{\nabla \psi}{n_1}\right|\leq \kappa,
\] 
then \eqref{eq:conditionondiscriminant} holds.
On the other hand, if
\[
\left| x-\dfrac{\nabla \psi}{n_1}\right|>\kappa
\] 
then \eqref{eq:conditionondiscriminant} holds when either
%\begin{dmath*}
%n_1^2(x \cdot \nu)^2 -2n_1(\nabla \psi \cdot \nu)(x\cdot \nu)+n_2^2-n_1^2+2n_1\nabla \psi \cdot x -|\nabla \psi|^2 +(\nabla \psi \cdot \nu)^2 \geq 0,
%\end{dmath*}
\begin{equation*}
 x \cdot \nu \geq \frac{\nabla \psi}{n_1} \cdot \nu +\sqrt{\left| x-\frac{\nabla \psi}{n_1}\right|^2 -\kappa^2} \quad \text{  or  } \quad x \cdot \nu \leq \frac{\nabla \psi}{n_1} \cdot \nu -\sqrt{\left| x-\frac{\nabla \psi}{n_1}\right|^2 -\kappa^2}.
\end{equation*}
Therefore, the critical angles between $x$ and $\nu$ are $\theta_c$ with
%\begin{equation*}
% x \cdot \nu=\cos \theta_c \geq \frac{\nabla \psi}{n_1} \cdot \nu +\sqrt{\left| x-\frac{\nabla \psi}{n_1}\right|^2 -\kappa^2} \quad \text{  or  } \quad x \cdot \nu \leq \frac{\nabla \psi}{n_1} \cdot \nu -\sqrt{\left| x-\frac{\nabla \psi}{n_1}\right|^2 -\kappa^2}.
%\end{equation*}
\begin{dmath*}
\begin{aligned}
& x \cdot \nu=\cos \theta_c = \frac{\nabla \psi}{n_1} \cdot \nu +\sqrt{\left| x-\frac{\nabla \psi}{n_1}\right|^2 -\kappa^2} \\
&\quad \text{or} \quad  \\ 
& x \cdot \nu =\cos \theta_c= \frac{\nabla \psi}{n_1} \cdot \nu -\sqrt{\left| x-\frac{\nabla \psi}{n_1}\right|^2 -\kappa^2}.
\end{aligned}
\end{dmath*}
%Reminding that we want $x\cdot \nu\geq 0$, the critical angles exist if and only if $\left| x-\frac{\nabla \psi}{n_1}\right|^2 -\kappa^2 \geq 0$, for
%$\left| x-\frac{\nabla \psi}{n_1}\right|^2 -\kappa^2 < 0$ we don't have critical angles.
\begin{remark}\rm
In two dimensions the critical angles are considered in  \cite{yu2011light}. 
It is assumed there that the interface $\Gamma$ is the $x$-axis, the region $y>0$ is filled with a material with refractive index $n_1$, and the region $y<0$ with a material with refractive index $n_2$. 
Also the phase discontinuity satisfies that $\nabla \psi$ is constant and is tangential to the interface, i.e., $\nabla \psi = (a,0)$ with, for example, $a>0$. 
Therefore, the above calculations applied to this case yield
\begin{align*}
\cos \theta_c=x \cdot \nu = \sqrt{\left| x-\frac{\nabla \psi}{n_1}\right|^2 -\kappa^2}=\sqrt{1-\frac{2|\nabla\psi|}{n_1}\cos(\pi/2 - \theta_c) + \frac{|\nabla\psi|^2}{n_1^2}-\kappa^2},
\end{align*} 
where $\kappa=\dfrac{n_2}{n_1}$.
Squaring both sides we obtain
\begin{dmath*}
\cos^2 \theta_c=1-\frac{2|\nabla\psi|}{n_1}\sin \theta_c + \frac{|\nabla\psi|^2}{n_1^2}-\kappa^2,
\end{dmath*} 
and the critical angles $\theta_c$ are therefore the solutions to the equation
\begin{dmath*}
\sin^2 \theta_c -\frac{2|\nabla\psi|}{n_1}\sin \theta_c + \frac{|\nabla\psi|^2}{n_1^2}-\kappa^2=0,
\end{dmath*} 
i.e.,
\begin{dmath*}
\theta_c=\arcsin \left( \frac{|\nabla\psi|}{n_1} \pm \kappa \right),
\end{dmath*}
which is in agreement with \cite[Formula (3)]{yu2011light}.

In three dimensions the critical angles are considered in \cite{aieta2012reflection}. 
The interface $\Gamma$ is the $xy$-plane, the region $z>0$ is filled with a material with refractive index $n_1$, and the region $z<0$ with a material with refractive index $n_2$. 
Also the phase discontinuity is tangential to the interface, i.e., $\nabla \psi = \left(\frac{\partial \psi}{\partial x},\frac{\partial \psi}{\partial y},0\right)$ and without loss of generality we may assume $x=(0,y,z)$. 
Once again, the above calculations applied to this case yield
\begin{align*}
\cos \theta_c=x \cdot \nu = \sqrt{\left| x-\frac{\nabla \psi}{n_1}\right|^2 -\kappa^2}=\sqrt{1-\frac{2}{n_1}\left|\frac{\partial \psi}{\partial y}\right|\cos(\pi/2 - \theta_c) + \frac{|\nabla\psi|^2}{n_1^2}-\kappa^2}.
\end{align*} 
%where $\kappa=\dfrac{n_2}{n_1}$.
Proceeding as before we find
%Squaring both sides we obtain
%\begin{dmath*}
%\cos^2 \theta_c=1--\frac{2}{n_1}\left|\frac{\partial \psi}{\partial y}\right|\sin \theta_c + \frac{|\nabla\psi|^2}{n_1^2}-\kappa^2,
%\end{dmath*} 
%$\theta_c$ are therefore the solutions to the equation
%\begin{dmath*}
%\sin^2 \theta_c --\frac{2}{n_1}\left|\frac{\partial \psi}{\partial y}\right|\sin \theta_c + \frac{|\nabla\psi|^2}{n_1^2}-\kappa^2=0,
%\end{dmath*} 
%i.e.,
\begin{dmath*}
\theta_c=\arcsin \left( \frac{1}{n_1}\frac{\partial \psi}{\partial y} \pm \sqrt{\kappa^2-\frac{1}{n_1^2}\left|\frac{\partial \psi}{\partial x} \right|^2 }\right),
\end{dmath*}
recovering \cite[Formula (8)]{aieta2012reflection}.
\end{remark}

\begin{remark}\rm
The reflection case is when $n_1=n_2$, so \eqref{vect2} and \eqref{lambdadependsonsign} become
\[
x-m=\dfrac{1}{n_1}\,\lambda\,\nu+\dfrac{\nabla \psi}{n_1}, \qquad 
\lambda=(n_1\,x - \nabla \psi ) \cdot \nu + \sqrt{ n_1^2 - \left( |n_1 x -\nabla \psi|^2 - \left[ (n_1 x -\nabla \psi)\cdot \nu \right]^2\right) },
\]
with $x$ the unit incident direction, $m$ the unit reflected vector, $\nu$ the unit normal to the interface at the striking point, and $\nabla \psi$ at the striking point.
Notice that the choice of the plus sign in front of the square root is because for reflection $m\cdot \nu\leq 0$.
\end{remark}

\section{Far field uniformly refracting planar and spherical metalenses}\label{sec:Far field uniformly refracting surfaces}
\setcounter{equation}{0}
%\subsection{The Far Field Case}
Let $\Gamma$ be a surface in three dimensional space and $V$ be a vector valued function defined on $\Gamma$; $V:\Gamma \to \R^3$.
If $x$ is an incident unit direction striking $\Gamma$ at a point $P$, and $m$ is the unit refracted direction, then we obtain, dividing by $n_1$ in
the generalized Snell law \eqref{vect2}, that
\begin{equation}\label{eq:generalizedSnell}
x-\kappa\,m=\lambda\,\nu(P) +V(P)
\end{equation}
where $\nu(P)$ is the unit outer normal to $\Gamma$ at $P$ for some $\lambda \in \R$; $\kappa=n_2/n_1$.

%We assume the refraction law in the following form
%\begin{equation}\label{eq:generalizedSnell}
%x-\kappa\,m=\lambda\,\nu + V(x)
%\footnote{Actually, the vector $V(x)$ depends on the point where the direction $x$ strikes the separation surface. In other words, the direction $x$ strikes the surface at a point $P$ and the normal $\nu$ is also at $P$. That is, $V$ should be a function of the incident point.}
%\end{equation}
%where $x$=incident direction, $m$=refracted direction, all unit vectors in $\R^3$, $\nu$ is the unit normal at the incidence point $P$,
%$\lambda$ is a scalar that can be calculated in terms of $x,\nu,\kappa,V$; and $V:S^2\to \R^3$ is a vector valued function. Suppose the function $V$ depends only on the incident direction $x$ (it might depend in principle from the point on the separation surface where the ray $x$ strikes it, perhaps this is nonsense since we do not know the surface a priori). As usual $\kappa=n_2/n_1$. 

Suppose rays emanate from the origin and we are given a fixed unit vector $m$.
Our goal is to answer the following two questions. First, given a surface $\Gamma$ 
separating media $n_1$ and $n_2$, find a field $V$ defined on $\Gamma$ so that all rays from the origin are refracted into the direction $m$. The second question is, given a field $V$ defined in a region of $\mathbb{R}^3$, find a separation surface $\Gamma$ between $n_1$ and $n_2$ within that region so that all rays emanating from the origin are refracted into the direction $m$.

We begin in this section answering the first question when $\Gamma$ is either a plane or a sphere, surfaces of traditional interest in optics, showing explicit phase discontinuities.
For general surfaces, the first question is considered in Section \ref{sec:m variable}, even for the more general case of variable $m$. The second question is answered in Section \ref{subsec:given the field find the phase discontinuity}.

\subsection{Case of the plane}\label{subsec:case of the plane}
\begin{figure}[htp]
\begin{center}
    \subfigure[$ $]{\label{fig:case of the plane}\includegraphics[width=2.9in]{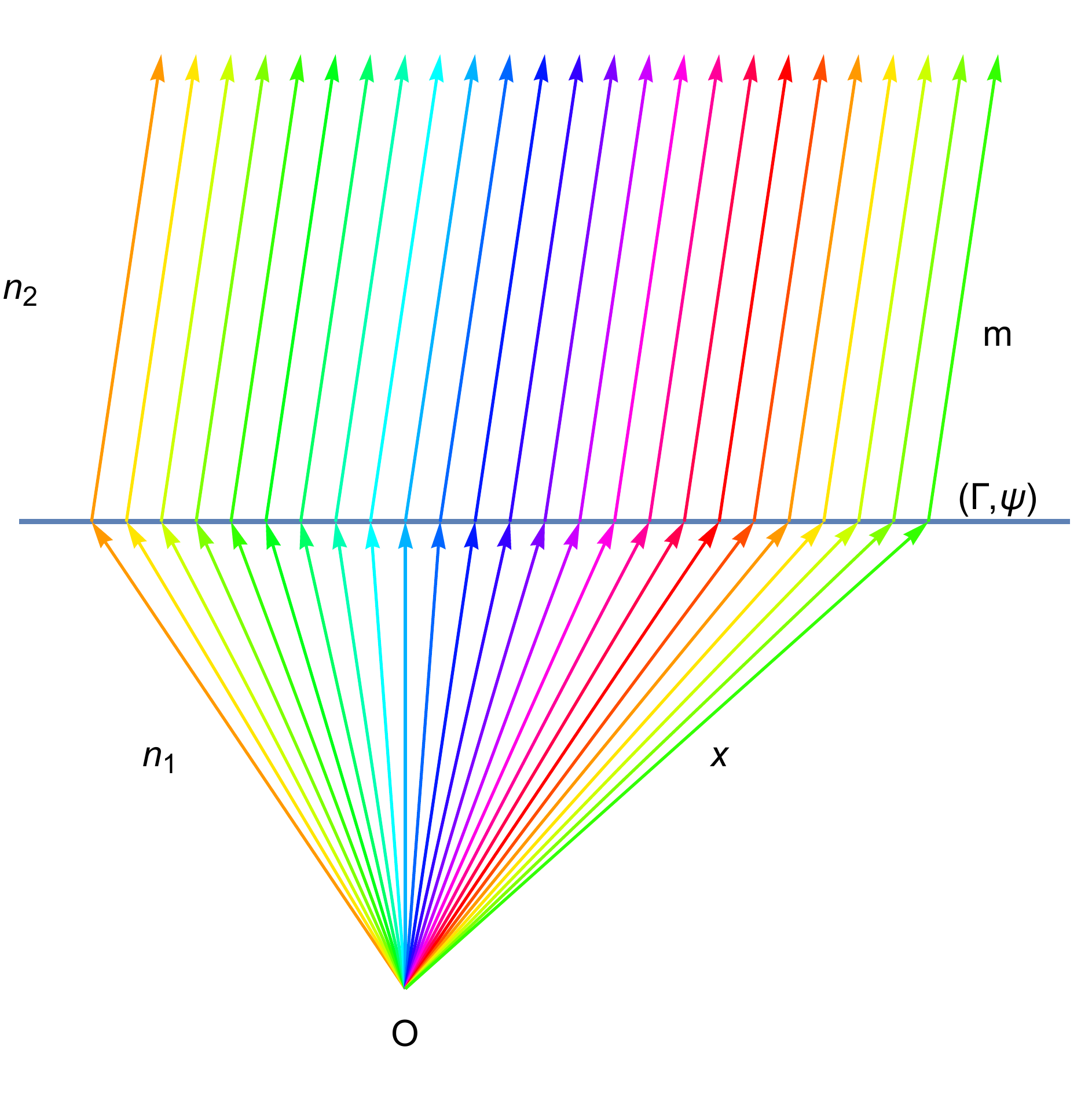}}
    \subfigure[$ $]{\label{fig:case of the sphere}\includegraphics[width=2.9in]{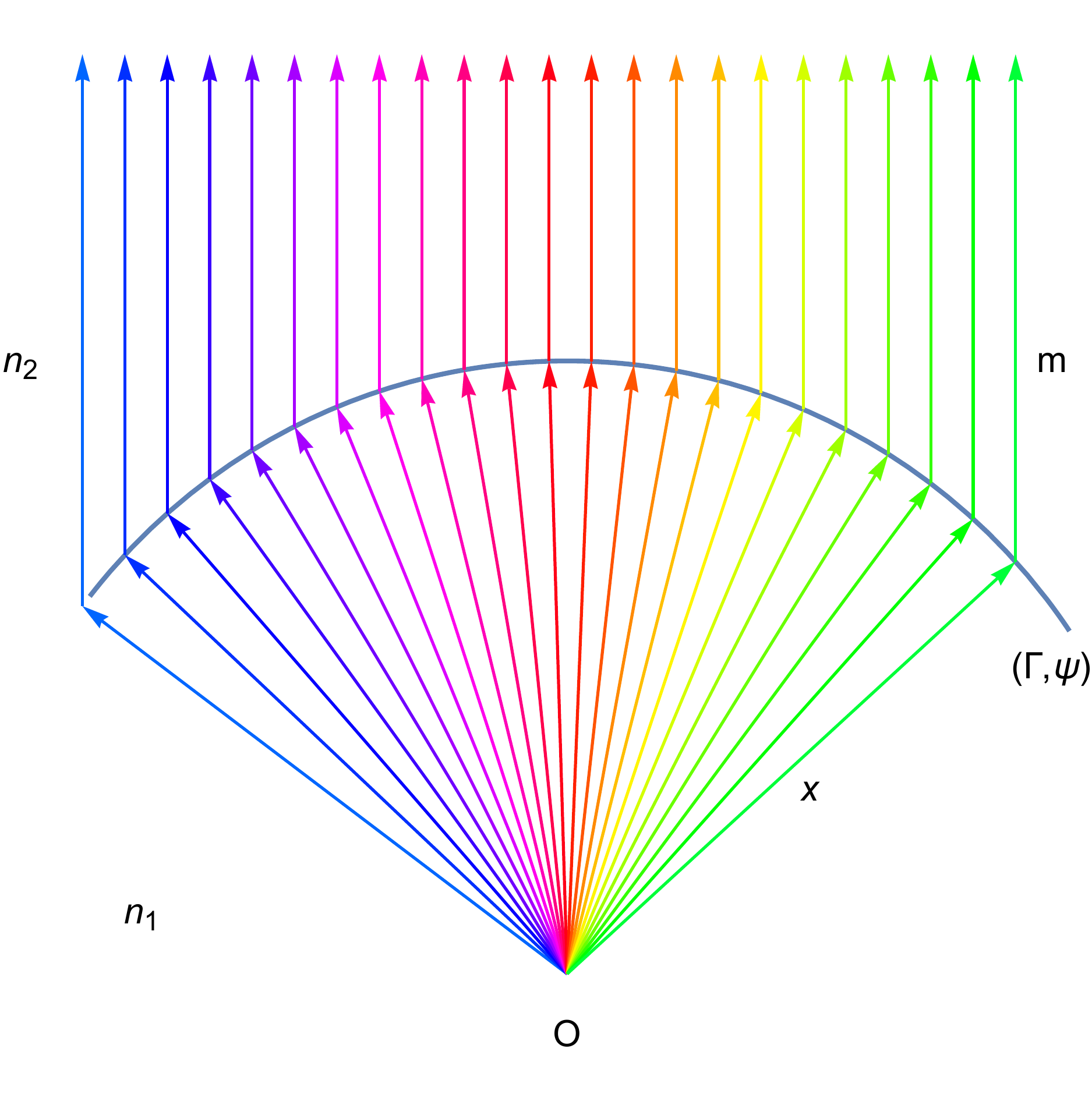}} 
%    \subfigure[After Sobel edge detection]{\label{fig:edge-c}\includegraphics[scale=1]{SeveralCartesianOvalskappa<1.pdf}}
\end{center}
  \caption{Planar and spherical metalenses}
  \label{fig:cases of plane and sphere}
\end{figure}

Let $\Gamma$ be the plane $x_1=a$ in $\mathbb{R}^3$ with $a>0$.
We want to determine a field $V=(V_1,V_2,V_3)$ defined on $\Gamma$ so that all rays emanating from the origin are refracted into the unit direction $m=(m_1,m_2,m_3)$, with $m_1>0$, Figure \ref{fig:case of the plane}. 

Using spherical coordinates $x(u,v)=(\cos u\,\sin v,\sin u\,\sin v,\cos v)$,
$0\leq u\leq 2\pi,0\leq v\leq \pi$, $\Gamma$ is described parametrically by 
\begin{equation}\label{eq:parametricequationofplane}
r(u,v)=\dfrac{a}{\cos u\,\sin v}\,x(u,v)=a\,\left(1,\tan u, \dfrac{1}{\cos u\,\tan v}\right).
\end{equation}
Since the normal to the plane $\Gamma$ is $\nu=(1,0,0)$, then \eqref{eq:generalizedSnell} implies that $\sin u\,\sin v -\kappa\,m_2=V_2(r(u,v))$ and $\cos v -\kappa\,m_3=V_3(r(u,v))$. 
Hence $V_2$ and $V_3$ are univocally determined. 
Also, from \eqref{eq:generalizedSnell} we  get 
\begin{equation}\label{eq:condition for V1}
V_1(r(u,v))=\cos u\,\sin v -\kappa\,m_1-\lambda(u,v).
\end{equation}
Notice also that from \eqref{lambdadependsonsign}, 
\[
\lambda=\nu\cdot(x-V)-\sqrt{\left(\nu\cdot (x-V) \right)^2-|x-V|^2+\kappa^2},
\]
which in the present case yields
\begin{align*}
\lambda&=\cos u\,\sin v-V_1-\sqrt{\kappa^2-(\sin u\,\sin v-V_2)^2-(\cos v-V_3)^2}\\
&=\cos u\,\sin v-V_1-\sqrt{\kappa^2-(\kappa\,m_2)^2-(\kappa\,m_3)^2}\\
&=\cos u\,\sin v-V_1-\kappa\,m_1\qquad \text{since $m_1>0$}\\
&=\dfrac{a}{\sqrt{a^2+x_2^2+x_3^2}}-V_1(a,x_2,x_3)-\kappa\,m_1.
\end{align*}
This means that in \eqref{eq:condition for V1} each $V_1$ determines $\lambda$ and vice-versa. 

We now write the field $V$ in rectangular coordinates $x_1,x_2,x_3$.
Since $\sqrt{a^2+x_2^2+x_3^2}= \dfrac{a}{\cos u\,\sin v}$, we can write
\begin{align*}
V_2(a,x_2,x_3)&=\dfrac{x_2}{\sqrt{a^2+x_2^2+x_3^2}}-\kappa\,m_2
=\left. \dfrac{\partial}{\partial x_2} \sqrt{x_1^2+x_2^2+x_3^2}\right|_{x_1=a}-\kappa\,m_2,\\
V_3(a,x_2,x_3)&=\dfrac{x_3}{\sqrt{a^2+x_2^2+x_3^2}}-\kappa\,m_3
=
\left. \dfrac{\partial}{\partial x_3} \sqrt{x_1^2+x_2^2+x_3^2}\right|_{x_1=a}-\kappa\,m_3\\
V_1(a,x_2,x_3)&=\dfrac{a}{\sqrt{a^2+x_2^2+x_3^2}}-\kappa\,m_1-\lambda
=
\left.\dfrac{\partial}{\partial x_1} \sqrt{x_1^2+x_2^2+x_3^2}\right|_{x_1=a}-\kappa\,m_1-\lambda, 
\end{align*}
for $-\infty<x_2,x_3<\infty$.
From \eqref{eq:parametricequationofplane}
$u=\arctan (x_2/a)$ and $v=\arctan\left(\dfrac{\sqrt{a^2+x_2^2}}{x_3} \right)$, so 
$\lambda(u,v)=h(x_2,x_3)$.
Let $\psi(x_1,x_2,x_3)=\sqrt{x_1^2+x_2^2+x_3^2}-\kappa\,m_1\,x_1-\kappa\,m_2\,x_2-\kappa\,m_3\,x_3$. 
%
%So $\Gamma$ has equation 
%\[
%r(t)=\dfrac{a}{\cos t}\left(\cos t, \sin t \right)=a\,(1,\tan t).
%\]
%From \eqref{eq:generalizedSnell}, $x-\kappa\,m- V(r(t))$ is a multiple of $\nu$ at $r(t)$, so
%\[
%r'(t)\cdot \left(x(t)-\kappa\,m- V(r(t)) \right)=0;\qquad x(t)=(\cos t, \sin t).
%\] 
%We have $r'(t)=(0,a\,\sec^2 t)$, so
%\[
%a\,\sec^2 t \,(\sin t-\kappa\,m_2-V_2(r(t)))=0.
%\]
%If $-\pi/2<t<\pi/2$, then this implies that $\sin t-\kappa\,m_2-V_2(r(t))=0$, that is,
%\[
%V_2(r(t))=\sin t -\kappa\,m_2,
%\]
%which gives 
%\[
%V_2(a,y)=\text{sign $y$}\,\sqrt{\dfrac{y^2}{a^2+y^2}}-\kappa\,m_2, \qquad -\infty<y<\infty.
%\]

%Notice that $V_1(r(u,v))$ determines $\lambda(r(u,v))$ and vice-versa, so the component $V_1$ can be picked up arbitrarily.
Therefore, if on the plane $x=a$ we give the field 
\begin{equation}\label{eq:fieldforplanex=a}
V(x_1,x_2,x_3):=\nabla \psi(x_1,x_2,x_3)-h(x_2,x_3)\,\mathbf{i},
\end{equation} 
then resulting metasurface does the desired refraction job.
If we want $V$ to be the gradient of a function, then $h(x_2,x_3)\,\mathbf{i}$ must be a gradient, which is only possible when $h(x_2,x_3)=C_0$ a constant;
that is, $V=\nabla \left(\psi(x_1,x_2,x_3)-C_0\,x_1 \right)$.
%Notice that, $V=\nabla \psi$, with $\psi(x_1,x_2,x_3)=\sqrt{x_1^2+x_2^2+x_3^2}-\kappa\,m_1\,x_1-\kappa\,m_2\,x_2-\kappa\,m_3\,x_3$. 
As a particular case when $m_1=1$, $m_2=m_3=0$, and $C_0=0$, we obtain the equivalent \cite[Formula (2)]{yu-capasso:flatoptics} (where a different orientation of the coordinates is used) with $x_1=a=f$. 
Notice also that if we want $V$ in \eqref{eq:fieldforplanex=a} to be tangential to the plane $x_1=a$, that is, $\left(\nabla \psi(a,x_2,x_3)-h(x_2,x_3)\,\mathbf{i}\right)\cdot (1,0,0)=0$, then $h=\dfrac{a}{\sqrt{a^2+x_2^2+x_3^2}}-\kappa\,m_1$.
 
\subsection{Case of the sphere}\label{subsec:caseofthespherecenteredatzero}
Now, the surface $\Gamma$ considered is a sphere of radius $R$ centered at the origin, that is, $r(u,v)=R\,x(u,v)$, with $x(u,v)$ spherical coordinates. We denote by $x=x(u,v)$; Figure \ref{fig:case of the sphere}. Since $\Gamma$ is a sphere, the normal $\nu=x$ and from \eqref{eq:generalizedSnell} we get
$
(x-\kappa\,m-V)\times x=0,
$
so
\begin{equation}\label{eq:conditionforVinthesphere}
\left(V+\kappa\,m\right)\times x=0.
\end{equation} 
That is, 
\begin{equation*}
\left[ 
\begin{matrix}
x_2 & -x_1 & 0\\
-x_3 & 0 & x_1\\
0 & x_3 &-x_2
\end{matrix}
\right]
\left( 
\begin{matrix}
V_1+\kappa\,m_1\\
V_2+\kappa\,m_2\\
V_3+\kappa\,m_3
\end{matrix}
\right)
=0.
\end{equation*}
Notice that $\det \left[ 
\begin{matrix}
x_2 & -x_1 & 0\\
-x_3 & 0 & x_1\\
0 & x_3 &-x_2
\end{matrix}
\right]=0$. Set $W_i=V_i+\kappa\,m_i$, so the system is equivalent to
\begin{equation*}
\left[ 
\begin{matrix}
0 & 0 & 0\\
x_2 x_3 & -x_1x_3 & 0\\
0 & x_1 x_3 &-x_1x_2
\end{matrix}
\right]
\left( 
\begin{matrix}
W_1\\
W_2\\
W_3
\end{matrix}
\right)
=0.
\end{equation*}
If $x_1x_2x_3\neq 0$, the last matrix has rank two, so the space of solutions has dimension one and the solutions are given by 
\[
(W_1,W_2,W_3)=\left(\dfrac{x_1}{x_3}, \dfrac{x_2}{x_3},1 \right)W_3,
\]
with $W_3$ arbitrary.
Therefore, 
\begin{align*}
V_1\left(R\,x(u,v)\right)&=\dfrac{x_1}{x_3}\left(V_3\left(R\,x(u,v)\right)+\kappa\,m_3\right)-\kappa\,m_1\\
V_2\left(R\,x(u,v)\right)&=\dfrac{x_2}{x_3}\left(V_3\left(R\,x(u,v)\right)+\kappa\,m_3\right)-\kappa\,m_2,
\end{align*}
with $V_3$ arbitrary.

Notice that if in \eqref{eq:conditionforVinthesphere} we take cross product with $x$,  we get
\begin{align*}
0&=x\times \left( \left(V+\kappa\,m\right)\times x\right)\\
&=
\left(V+\kappa\,m\right)\,(x\cdot x)- x\, \left( \left(V+\kappa\,m\right)\cdot x\right)\\
&=
V+\kappa\,m-\left( \kappa \,(m\cdot x)+V\cdot x\right)\,x.
\end{align*}
Hence, if 
we want to pick $V$ tangential to the sphere, we obtain
\[
V(R\,x)=-\kappa\,m+\kappa\,(m\cdot x)\,x\text{ with $|x|=1$}.
\]
$V$ is a field defined on the sphere of radius $R$.
We shall determine a function $\psi$ defined in a neighborhood of the sphere of radius $R$ such that $V(Rx)=\left.\nabla\psi(Rx)\right|_{|x|=1}$, and satisfying 
\begin{equation}\label{eq:equationforgradpsionthesphere}
\psi_{x_j}(R\,x)=-\kappa\,m_j+\kappa\,(m\cdot x)\,x_j, \text{ for $|x|=1$}, \quad 1\leq j\leq 3.
\end{equation}
In fact, we have ($x=x(u,v)$)
\begin{align*}
\dfrac{\partial \psi(Rx(u,v))}{\partial u}&=R\, \sum_{k=1}^3\dfrac{\partial \psi}{\partial x_k}(Rx(u,v))\,(x_k)_u=R \,(D\psi)(Rx(u,v))\cdot x_u\\
&=
R\,\left(-\kappa\,m\cdot x_u+\kappa\,(m\cdot x)\,(x\cdot x_u) \right)=
-\kappa\,R\,(m\cdot x_u)=-\kappa\,R\,\dfrac{\partial }{\partial u}(m\cdot x),
\end{align*}
and similarly,
\begin{align*}
\dfrac{\partial \psi(Rx(u,v))}{\partial v}&=-\kappa\,R\,\dfrac{\partial }{\partial v}(m\cdot x).
\end{align*}
Integrating the derivative in $u$ yields
\[
\psi(Rx(u,v))=-\kappa\,R\,(m\cdot x)+g(v),
\]
and integrating the derivate in $v$ we obtain
\[
\psi(Rx(u,v))=-\kappa\,R\,(m\cdot x(u,v))+C_1,
\]
with $C_1$ an arbitrary constant.
Writing this in rectangular coordinates yields
\[
\psi\left(R\,(z_1,z_2,z_3)\right)=-\kappa\,R\,(m\cdot (z_1,z_2,z_3))+C_1,\text{ for $|(z_1,z_2,z_3)|=1$.}
\]
We now define $\psi$ on a neighborhood of $|z|=R$ so that \eqref{eq:equationforgradpsionthesphere} holds.
Define 
\begin{equation}\label{eq:phasediscontinuityforsphericalcase}
\psi(z)=-\kappa\,R\,(m\cdot z)\,|z|^{-1}+C_1,\text{ for $R-\epsilon<|z|<R+\epsilon$}.
\end{equation}
We have
\[
\nabla \psi(z)=-\kappa\,R\,m\,|z|^{-1}+ \kappa\,R\, (m\cdot z) \,z\,|z|^{-3},
\]
so for $z=Rx$,  with $|x|=1$, we obtain
\[
\nabla \psi(Rx)=-\kappa\,m+ \kappa\,(m\cdot x)\,x
\]
as desired. 
Therefore the phase discontinuity $\psi$ from \eqref{eq:phasediscontinuityforsphericalcase} has gradient tangential to the sphere and can be placed on the spherical interface $|z|=R$ so that all rays from the origin are refracted into the fixed direction $m$.

\section{Metalenses refracting into a set of variable directions}\label{sec:m variable}
\setcounter{equation}{0}
Suppose $m(u,v)=(m_1(u,v),m_2(u,v),m_3(u,v))$ is a given $C^2$ unit field of directions,
and let $\Gamma$ be a $C^2$ surface given parametrically by $r(u,v)=\rho(u,v)\,x(u,v)$ where $x(u,v)$ are spherical coordinates and $\rho(u,v)>0$ is the polar radius.
We want to see when is it possible to have a phase discontinuity $\psi$ on the surface $\Gamma$ so that each ray from the origin with direction $x(u,v)$ is refracted into the  direction $m(u,v)$. 
%Assume $\rho(u,v)$ is so that $r(u,v)$ describes a sphere centered at $(0,0,a)$ with radius $R<a$. 
%The normal $\nu=\dfrac{(0,0,a)-r(u,v)}{|(0,0,a)-r(u,v)|}$, so
From \eqref{eq:generalizedSnell}
\[
x(u,v)-\kappa\,m(u,v)-V(r(u,v))=\lambda\,\nu(r(u,v))
\]
so
\[
\left(x-\kappa\,m-V \right)\times \nu=0. 
\]
Taking cross product with $\nu$ yields
\[
0=\nu\times \left(\left(x-\kappa\,m-V \right)\times \nu \right)
=
\left(x-\kappa\,m-V \right)(\nu\cdot \nu)-\nu\,\left(\left(x-\kappa\,m-V \right)\cdot \nu\right).
\]
If $V$ is tangential to $\Gamma$, then $V\cdot \nu=0$ and so
\[
0=x-\kappa\,m-V-\left((x-\kappa\,m)\cdot \nu\right) \,\nu,
\]
that is,
\[
V=x-\kappa\,m-\left((x-\kappa\,m)\cdot \nu\right) \,\nu.
\]
If $V(r(u,v))=(\nabla \psi)(r(u,v))$, then
\[
\psi_{x_j}(r(u,v))=x_j(u,v)-\kappa\,m_j(u,v)-\left((x(u,v)-\kappa\,m(u,v))\cdot \nu(r(u,v))\right) \,\nu_j(r(u,v)).
\]
Since $\nu\cdot r_u=\nu\cdot r_v=0$ and $x\cdot x_u=x\cdot x_v=0$,
\begin{align*}
&\dfrac{\partial }{\partial u}\left( \psi(r(u,v))\right)\\
&=
(\nabla \psi)(r(u,v))\cdot r_u
=
(x-\kappa\,m)\cdot r_u-\left((x-\kappa\,m)\cdot \nu\right) \,(\nu\cdot r_u)\\
&=
(x-\kappa\,m)\cdot r_u=(x-\kappa\,m)\cdot \left(\rho_u\,x+\rho\,x_u\right)\\
&=\rho_u\,(x-\kappa\,m)\cdot x+ \rho\,(x-\kappa\,m)\cdot x_u\\
&=
\rho_u\,(1-\kappa\,m\cdot x)-\kappa\, \rho\,m\cdot x_u=
\rho_u\,(1-\kappa\,m\cdot x)-\kappa\, \rho\,(m\cdot x)_u+\kappa\, \rho\,(m_u\cdot x)\\
&=\left\{\rho\,(1-\kappa\,m\cdot x) \right\}_u+\kappa\, \rho\,(m_u\cdot x),
\end{align*}
and similarly 
\begin{align*}
\dfrac{\partial }{\partial v}\left( \psi(r(u,v))\right)
&=
\left\{\rho\,(1-\kappa\,m\cdot x) \right\}_v+\kappa\, \rho\,(m_v\cdot x).
\end{align*}
Let us now consider the first order system in $\Phi$
\begin{equation}\label{eq:system for Phi general m}
\begin{cases}
\Phi_u&=\kappa\, \rho\,(m_u\cdot x)\\
 \Phi_v&=\kappa\, \rho\,(m_v\cdot x).
 \end{cases}
\end{equation}
where $\Phi(u,v) = \psi(r(u,v))-\rho\;(1-\kappa m \cdot x)$. 
If the given set of directions $m(u,v)$ and the surface $\Gamma$ satisfy
\begin{equation}\label{eq:compatibility condition between m and Gamma}
m_u\cdot r_v=m_v\cdot r_u,
\end{equation}
then by  \cite[Chapter 6, pp. 117-118]{hartman-book-odes}(see also \eqref{eq:Hartmanconditionforsolvability} below) there exists $\Phi$ solving \eqref{eq:system for Phi general m}.
By integration we then obtain that the phase discontinuity $\psi$ satisfies along $\Gamma$ that
\begin{equation}\label{eq:valuesofpsionthesurfacegeneralm}
\psi(r(u,v))=\rho\,(1-\kappa\,m\cdot x)+\Phi(u,v)=|r(u,v)|-\kappa\,(m(u,v)\cdot r(u,v))+\Phi(u,v).
\end{equation}
To find the gradient of $\psi$ we need to have $\psi$ defined in a neighborhood of the surface $r(u,v)$ such that \eqref{eq:valuesofpsionthesurfacegeneralm} holds and that its gradient satisfies on $r(u,v)$ 
\begin{equation}\label{eq:gradientconditiononthesurfacegeneralm}
(\nabla\psi)(r(u,v))=x-\kappa\,m-\left((x-\kappa\,m)\cdot \nu\right) \,\nu.
\end{equation}
Notice that this implies $(\nabla \psi)(r(u,v))\perp \nu$.
To construct the function $\psi$ in a neighborhood of the surface $\Gamma$ (we will construct it in a neighborhood of each point in $\Gamma$), given parametrically by $r(u,v)$, we use the notion of envelope from classical differential geometry; see for example \cite[Chapter 5, Section 4]{Pogorelov-diff-geometry} or \cite[Chapter 3]{docarmo:differential-geometry-of-surfaces}. We will actually construct a surface that is developable, in particular, it has Gauss curvature zero.
For a recent reference on developable surfaces, its applications and design see \cite{Tang:2016:IDD:2882845.2832906}. 
%\url{http://www.geometrie.tugraz.at/wallner/abw.pdf}.

Since the required $\psi$ must satisfy \eqref{eq:valuesofpsionthesurfacegeneralm}, consider the surface $\Gamma'$ given parametrically by 
\begin{equation}\label{eq:new vector P with Phi}
P(u,v)=\left( r(u,v),|r(u,v)|-\kappa\,(m(u,v)\cdot r(u,v))+\Phi(u,v)\right)
\end{equation} 
in four dimensions. At each point $P(u,v)$, consider the 4-dimensional vector 
\[
N(u,v)=\left(x-\kappa\,m-\left((x-\kappa\,m)\cdot \nu\right) \,\nu,-1 \right),
%:=\left(N^*(u,v),-1 \right),
\] 
where $x=x(u,v)$ and $\nu$ is the unit normal to the surface $\Gamma$ at $r(u,v)$.
Next consider the plane $\Pi_{uv}$ passing through the point $P(u,v)$ and with normal $N(u,v)$, that is, in coordinates $x_1,x_2,x_3,x_4$, $\Pi_{uv}$ has equation 
\begin{equation}\label{eq:formulafortheplaneonthecurvegeneralm}
F(x_1,x_2,x_3,x_4,u,v):=N(u,v)\cdot \left( (x_1,x_2,x_3,x_4)-P(u,v)\right)=0.
\end{equation}
%We now repeat the argument in the previous section with this function $F$ leading to the system
%\eqref{eq:systemforFuandFuuFvvFuv} with $P$ given by \eqref{eq:new vector P with Phi}.
%Now \eqref{eq:NdotPuandPvarezero} holds with this new $P$ because writing

Therefore we have a family of planes $\Pi_{uv}$ depending on the parameters $u,v$, and
we will let $x_4=\psi(x_1,x_2,x_3)$ be by definition the envelope to this family of planes.
Of course, we need to know under what conditions on $r(u,v)$ and $m(u,v)$ this envelope $\psi$ exists. It will be defined by solving the system of equations
\begin{align}\label{eq:systemofthreeequationsforimplicitfunctionmvar}
\begin{cases}
F(x_1,x_2,x_3,x_4,u,v)&=0\\[0.5ex]
\dfrac{\partial F}{\partial u}(x_1,x_2,x_3,x_4,u,v)&=0\\[2ex]
\dfrac{\partial F}{\partial v}(x_1,x_2,x_3,x_4,u,v)&=0.
\end{cases}
\end{align}
In fact, let us fix values $u=u_0,v=v_0$, and let $P_0=P(u_0,v_0)=(p_1,p_2,p_3,p_4)$ be the corresponding value on the surface $\Gamma'$; 
and consider the map 
\[
G\left(x_1,x_2,x_3,x_4,u,v\right)=\left(F(x_1,x_2,x_3,x_4,u,v), \dfrac{\partial F}{\partial u}(x_1,x_2,x_3,x_4,u,v),\dfrac{\partial F}{\partial v}(x_1,x_2,x_3,x_4,u,v) \right).
\]
The function $G$ has continuous partial derivatives in a neighborhood of the point $\left(p_1,p_2,p_3,p_4,u_0,v_0\right)$, and 
\[
G\left(p_1,p_2,p_3,p_4,u_0,v_0\right)=0.
\]
By the implicit function theorem, if the Jacobian determinant
\begin{equation}\label{eq:conditionfortheimplicitfunctiontheoremmvar}
\dfrac{\partial G}{\partial (x_4, u,v) }\left(p_1,p_2,p_3,p_4,u_0,v_0 \right)
=
\left.\det
\left(
\begin{matrix}
\dfrac{\partial F}{\partial x_4} & \dfrac{\partial F}{\partial u}& \dfrac{\partial F}{ \partial v}\\[2ex]
\dfrac{\partial^2 F}{\partial x_4\partial u} & \dfrac{\partial^2 F}{\partial u\partial u}& \dfrac{\partial^2 F}{ \partial v\partial u}\\[2ex]
\dfrac{\partial^2 F}{\partial x_4\partial v} & \dfrac{\partial^2 F}{\partial u\partial v}& \dfrac{\partial^2 F}{ \partial v\partial v}
\end{matrix} \right)\right|_{\left(p_1,p_2,p_3,p_4,u_0,v_0 \right)}\neq 0,
\end{equation}
then there are unique differentiable functions $g_1,g_2,g_3$ in the variables $x_1,x_2,x_3$ defined in a neighborhood $U$ of $(p_1,p_2,p_3)$ such that $p_4=g_1(p_1,p_2,p_3)$, $u_0=g_2(p_1,p_2,p_3)$ and $v_0=g_3(p_1,p_2,p_3)$ with
\[
G\left(x_1,x_2,x_3,g_1(x_1,x_2,x_3),g_2(x_1,x_2,x_3),g_3(x_1,x_2,x_3) \right)=0
\]
for all $(x_1,x_2,x_3)\in U$. Therefore, if we let $\psi(x_1,x_2,x_3)=g_1(x_1,x_2,x_3)$ for $(x_1,x_2,x_3)\in U$, then $\psi$ is the function we need, i.e., $\psi$ is by construction defined in a neighborhood of the point $(p_1,p_2,p_3)\in \Gamma$ and satisfies \eqref{eq:valuesofpsionthesurfacegeneralm} and \eqref{eq:gradientconditiononthesurfacegeneralm}.
%We also have from \eqref{eq:formulafortheplaneonthecurvegeneralm} that\marginpar{not needed}
%\[
%F(x_1,x_2,x_3,x_4,u,v)=\left((x_1,x_2,x_3)- r(u,v) \right)\cdot N^*(u,v)-\left(x_4- |r(u,v)|-\kappa\,(m\cdot r(u,v))+\Phi(u,v) \right)
%\]
%and hence
%\begin{align*}
%g_1(x_1,x_2,x_3)&=\left((x_1,x_2,x_3)- r(g_2(x_1,x_2,x_3),g_3(x_1,x_2,x_3)) \right)\cdot N^*(g_2(x_1,x_2,x_3),g_3(x_1,x_2,x_3))\\
%&\qquad + |r(g_2(x_1,x_2,x_3),g_3(x_1,x_2,x_3))|+\kappa\,(m\cdot r(g_2(x_1,x_2,x_3),g_3(x_1,x_2,x_3)))\\
%&\qquad -\Phi(g_2(x_1,x_2,x_3),g_3(x_1,x_2,x_3)). 
%\end{align*}

We now analyze under what conditions on the surface $\Gamma$ and $m$, \eqref{eq:conditionfortheimplicitfunctiontheoremmvar} holds.
Notice first that since $\partial_{x_4}F=-1$, the matrix inside the determinant in \eqref{eq:conditionfortheimplicitfunctiontheoremmvar} equals
\[
\left(
\begin{matrix}
1 & \dfrac{\partial F}{\partial u}& \dfrac{\partial F}{ \partial v}\\[2ex]
0 & \dfrac{\partial^2 F}{\partial u\partial u}& \dfrac{\partial^2 F}{ \partial v\partial u}\\[2ex]
0 & \dfrac{\partial^2 F}{\partial u\partial v}& \dfrac{\partial^2 F}{ \partial v\partial v}
\end{matrix} \right),
\]
and therefore \eqref{eq:conditionfortheimplicitfunctiontheoremmvar} means
\[
\det \left(
\begin{matrix}
\dfrac{\partial^2 F}{\partial u\partial u}& \dfrac{\partial^2 F}{ \partial v\partial u}\\[2ex]
\dfrac{\partial^2 F}{\partial u\partial v}& \dfrac{\partial^2 F}{ \partial v\partial v}
\end{matrix} \right)\neq 0.
\]
Let us find what this means in terms of the initial surface $\Gamma$ and the field $m$.
To simplify the notation let $X=(x_1,x_2,x_3,x_4)$, so we can write \eqref{eq:formulafortheplaneonthecurvegeneralm} as
\[
F(X,u,v)=N(u,v)\cdot \left( X-P(u,v)\right).
\]
By calculation
\begin{align}\label{eq:systemforFuandFuuFvvFuvmvar}
\begin{cases}
F_u&=N_u\cdot (X-P)-N\cdot P_u\\
F_{uu}&=N_{uu}\cdot (X-P)-2\,N_u\cdot P_u-N\cdot P_{uu}\\
F_{uv}&=N_{uv}\cdot (X-P)-N_u\cdot P_v-N_v\cdot P_u-N\cdot P_{uv}\\
F_{vv}&=N_{vv}\cdot (X-P)-2\,N_v\cdot P_v-N\cdot P_{vv}.
\end{cases}
\end{align}
We first show that 
\begin{equation}\label{eq:NdotPuandPvarezeromvar}
N\cdot P_u=N\cdot P_v=0.
\end{equation}
Indeed, we have

\[
P(u,v)
=
\rho(u,v)\,\left(x,1-\kappa\,m\cdot x \right)+(0,\Phi),
\]
so
%that $P_u$ and $P_v$ satisfy
% \eqref{eq:derivatives of P with u and v} since 
%$\Phi$ satisfies \eqref{eq:system for Phi general m}.
%Therefore as before we obtain \eqref{eq:NdotPuandPvarezero}. 

\begin{align}\label{eq:derivatives of P with u and v m variable}
P_u&=\rho_u\,\left(x,1-\kappa\,m\cdot x \right)+ \rho\,\left(x_u,-\kappa\,m\cdot x_u -\kappa\,m_u\cdot x\right)+(0,\Phi_u)\\
P_v&=\rho_v\,\left(x,1-\kappa\,m\cdot x \right)+ \rho\,\left(x_v,-\kappa\,m\cdot x_v -\kappa\,m_v\cdot x\right)+(0,\Phi_v).\notag
\end{align}
Hence
\begin{align*}
N\cdot P_u&=
\left\{\rho_u\,\left(x,1-\kappa\,m\cdot x \right)+ \rho\,\left(x_u,-\kappa\,m\cdot x_u -\kappa\,m_u\cdot x\right)+(0,\Phi_u)\right\}\cdot \\
&\qquad \qquad 
\left(x-\kappa\,m-\left((x-\kappa\,m)\cdot \nu\right) \,\nu,-1 \right)\\
&=\left(\rho_u\,x+ \rho\,x_u\right)\cdot 
\left(x-\kappa\,m-\left((x-\kappa\,m)\cdot \nu\right)\nu\right)
-
\rho_u\,\left(1-\kappa\,m\cdot x \right)+\\
&\qquad \qquad 
 \rho\,(\,\kappa\,m\cdot x_u+\kappa\,m_u\cdot x)-\Phi_u \\
&=
\rho_u -\rho_u\, \kappa\,x\cdot m -\rho\,\kappa\,x_u\cdot m
-
\rho_u\,+ \rho_u\,\kappa\,m\cdot x +\rho\,\kappa\,m\cdot x_u +\rho\,\kappa\,m_u\cdot x-\rho\,\kappa\,m_u\cdot x\\
&=0,
\end{align*}
since $\left(\rho_u\,x+ \rho\,x_u\right)\cdot \nu=r_u\cdot \nu=0$ and $x_u\cdot x=0$.
The same calculation with $P_v$ instead of $P_u$ yields the second identity in \eqref{eq:NdotPuandPvarezeromvar}.

Next, differentiating \eqref{eq:NdotPuandPvarezeromvar} with respect to $u$ and $v$ yields 
\[
N\cdot P_{uu}=-N_u\cdot P_u,\qquad N\cdot P_{uv}=-N_u\cdot P_v=-N_v\cdot P_u,\qquad
N\cdot P_{vv}=-N_v\cdot P_v,
\]
since $P_{uv}=P_{vu}$.
Hence letting $X=P$ in \eqref{eq:derivatives of P with u and v m variable} yields
\[
F_{uu}=-N_u\cdot P_u,\qquad F_{uv}=-N_v\cdot P_u=-N_u\cdot P_v,\qquad
F_{vv}=-N_v\cdot P_v.
\]
Now let us calculate these dot products. First set 
\[
B=(x-\kappa\,m)\cdot \nu
\]
and write
\begin{align*}
&N_u\cdot P_u\\
&=
\left\{\rho_u\,\left(x,1-\kappa\,m\cdot x \right)+ \rho\,\left(x_u,-\kappa\,m\cdot x_u \right)+(0,\Phi_u)\right\}\cdot\\
&\qquad \qquad
\left\{x_u-\kappa\, m_u-\left[(x-\kappa\,m)\cdot \nu\right]_u\,\nu-\left[(x-\kappa\,m)\cdot \nu\right]\,\nu_u,0 \right\}\\
&=
\left(\rho_u\,x+\rho\,x_u \right)
\cdot
\left(x_u-\kappa\,m_u-B_u\,\nu-B\,\nu_u \right)\\
&=
\left(\rho_u\,x+\rho\,x_u \right)
\cdot x_u
-
\kappa\,\left(\rho_u\,x+\rho\,x_u \right)
\cdot m_u
-
B_u\,\left(\rho_u\,x+\rho\,x_u \right)
\cdot
\nu
-
B\,\left(\rho_u\,x+\rho\,x_u \right)
\cdot
\nu_u\\
&=
\left(\sin^2v\right)\,\rho - \kappa\,\left(\rho_u\,x+\rho\,x_u \right)
\cdot m_u-
B\,\left(\rho_u\,x+\rho\,x_u \right)
\cdot
\nu_u=\left(\sin^2v\right)\,\rho - \kappa\,r_u\cdot m_u-
B\,r_u
\cdot
\nu_u,
\end{align*}
since $x\cdot x_u=0,x_u\cdot x_u=\sin^2v$, and $\left(\rho_u\,x+ \rho\,x_u\right)\cdot \nu=r_u\cdot \nu=0$.
Also $x_v\cdot x_v=1$ and $x_u\cdot x_v=0$, so we obtain similarly
\[
N_v\cdot P_v=\rho - \kappa\,r_v\cdot m_v-
B\,r_v
\cdot
\nu_v,\qquad
N_u\cdot P_v= - \kappa\,r_u\cdot m_v-
B\,r_u
\cdot
\nu_v.
\]
Next, differentiating $r_u\cdot \nu=r_v\cdot \nu=0$ yields 
\begin{equation}\label{eq:formulasfordotprodbetweenruandnuvmvar}
r_u\cdot \nu_u=-r_{uu}\cdot \nu,
\qquad
r_u\cdot \nu_v=-r_{uv}\cdot \nu,\qquad
r_v\cdot \nu_v=-r_{vv}\cdot \nu.
\end{equation} 
Therefore
\[
\left(
\begin{matrix}
F_{uu} & F_{uv}\\
F_{vu}& F_{vv}
\end{matrix} \right)
=
\left(
\begin{matrix}
-\left(\sin^2 v\right)\, \rho+ \kappa\,r_u\cdot m_u-B\,r_{uu}\cdot \nu &  \kappa\,r_u\cdot m_v-B\,r_{uv}\cdot \nu\\
\kappa\,r_u\cdot m_v-B\,r_{uv}\cdot \nu& -\rho+ \kappa\,r_v\cdot m_v -B\,r_{vv}\cdot \nu
\end{matrix} \right),
\]
and so
\begin{align}\label{formulafor2x2determinantmvar}
\det \left(
\begin{matrix}
F_{uu} & F_{uv}\\
F_{vu}& F_{vv}
\end{matrix} \right)
&=
%\left(\sin^2 v\right)\,\rho^2  -\left(\sin^2 v\right)\,\rho\,\kappa\,r_v\cdot m_v+\left(\sin^2 v\right)\,\rho\,B\,r_{vv}\cdot\nu \\
%&-\rho\,\kappa\,r_u\cdot\, m_u-B\,\kappa\,(r_u\cdot m_u)(r_{vv}\cdot \nu)+\kappa^2\,(r_u\cdot m_u)(r_v\cdot m_v)\nonumber\\
%&+\rho\,B\,r_{uu}\cdot \nu - B\,\kappa\,(r_{uu}\cdot \nu)(r_v\cdot m_v)-\kappa^2\,(r_u\cdot m_v)^2\nonumber\\
%&+2B\,\kappa\, (r_u\cdot m_v)(r_{uv}\cdot \nu)
%+
%B^2\,\det \left(
%\begin{matrix}
%r_{uu}\cdot \nu & r_{uv}\cdot \nu\\
%r_{uv}\cdot \nu& r_{vv}\cdot \nu
%\end{matrix} \right),\nonumber
\det \left[ -\rho\, \left(
\begin{matrix}
 x_u\cdot x_u & x_u\cdot x_v\\
 x_v\cdot x_u& x_v\cdot x_v
\end{matrix}\right)  
+\kappa\,\left(
\begin{matrix}
r_{u}\cdot m_u &r_{u}\cdot m_v\\
r_{v}\cdot m_u&r_{v}\cdot m_v
\end{matrix}\right)
-
B\,\left(
\begin{matrix}
r_{uu}\cdot\nu &r_{uv}\cdot \nu\\
r_{vu}\cdot\nu&r_{vv}\cdot \nu
\end{matrix}\right) \right],
\end{align}
with $B=(x-\kappa\,m)\cdot \nu$.
Notice that the first and third matrices in the last determinant are respectively the first fundamental
form of the 2-sphere, and the second fundamental form of the surface $\Gamma$.
%Notice that when $m$ is constant, the last equation yields \eqref{eq:formulafor2x2determinant}.

% (the matrices in the determinant are respectively the first fundamental
%form of the 2-sphere, the second fundamental form of the surface $\Gamma$ and $\nabla r \cdot (\nabla m)^t$).
%Notice that the Gauss curvature of $\Gamma$ equals
%\[
%K(u,v)=\dfrac{\det \left(
%\begin{matrix}
%r_{uu}\cdot \nu & r_{uv}\cdot \nu\\
%r_{uv}\cdot \nu& r_{vv}\cdot \nu
%\end{matrix} \right)}{\det \left(
%\begin{matrix}
%r_u\cdot r_u & r_u\cdot r_v\\
%r_u\cdot r_v & r_v\cdot r_v
%\end{matrix} \right)}
%\]
%and by calculation
%\[
%\det \left(
%\begin{matrix}
%r_u\cdot r_u & r_u\cdot r_v\\
%r_u\cdot r_v & r_v\cdot r_v
%\end{matrix} \right)=\rho^2\left( (\rho_u)^2+(\sin^2 v)\,\left((\rho_v)^2+\rho^2 \right) \right).
%\]
%So the determinant on the right hand side of \ref{formulafor2x2determinantmvar}
%equals
%\[
%K(u,v)\,\rho^2\left( (\rho_u)^2+(\sin^2 v)\,\left((\rho_v)^2+\rho^2 \right) \right).
%\]

{\it Therefore, we have proved the following:
if a variable field $m$ and a surface $\Gamma$ satisfy the compatibility condition \eqref{eq:compatibility condition between m and Gamma}, and  
the determinant \eqref{formulafor2x2determinantmvar} is not zero at a point $(u_0,v_0)$, then there is a neighborhood $U$ of the point $r(u_0,v_0)$ and a phase discontinuity function $\psi$ defined in $U$ for the surface $\Gamma$, with gradient $\nabla \psi$ tangential to $\Gamma$, so that it yields the desired refraction job, i.e., each ray emanating in the direction $x(u,v)$, for $(u,v)$ in a neighborhood of $(u_0,v_0)$, is refracted by the metasurface $(\Gamma,\psi)$ into the direction $m(u,v)$.}

\begin{remark}[Case when $m$ is a constant vector]\label{rmk:case when m is constant}\rm
If $m(u,v)=(m_1,m_2,m_3)$ is constant, then \eqref{eq:compatibility condition between m and Gamma} is clearly satisfied by any $\Gamma$ and in condition  \eqref{formulafor2x2determinantmvar} the second matrix on the right hand side is zero.\end{remark}

\begin{remark}\rm
To illustrate the determinant condition \eqref{formulafor2x2determinantmvar}, let us consider the special case when $\Gamma$ is a sphere centered at the origin, and $m$ is a constant vector. We have $r(u,v)=R\,x(u,v)$, and
$\nu=x(u,v)$. So $r_{uu}=Rx_{uu}$ and similarly for $r_{vv}$ and $r_{uv}$. Also
$B=1-\kappa\,m\cdot x$,  
$
x_{uu}\cdot x= -\sin^2 v, x_{uv}\cdot x=0$, and $x_{vv}\cdot x=-1.
$
Hence
$
r_{uu}\cdot x=-R\,\sin^2 v$,
$r_{uv}\cdot x=0$, and
$
r_{vv}\cdot x=-R.
$
Therefore the determinant in \eqref{formulafor2x2determinantmvar} equals
\begin{align*}
%&R^2\,\sin^2v +B\,R\left(-R\,\sin^2 v -R\,\sin^2 v\right)
%+
%B^2\,R^2\,\sin^2 v\\
%&=
%R^2\,\sin^2 v\left(1-2\,B +B^2\right)\\
%&=
R^2\,\sin^2 v\left(1-B\right)^2=
R^2\,\kappa^2 \,\left(\sin^2 v\right)\,(m\cdot x)^2.
\end{align*}
For example, if $m=(0,0,1)$, i.e., all rays are refracted vertically, then the determinant equals 
\[
R^2\,\kappa^2 \,\left(\sin v\,\cos v\right)^2=
\dfrac{R^2\,\kappa^2}{4} \,\sin^2 (2v)
\]
which is not zero as long as $v\neq \pi/2$ or zero.
This shows also that for the sphere the phase discontinuity $\psi$ exists and can be obtained by solving the system of equations 
%\eqref{eq:systemofthreeequationsforimplicitfunction}
\eqref{eq:systemofthreeequationsforimplicitfunctionmvar}. 
Notice that in this case a phase discontinuity $\psi$ was calculated explicitly in Section \ref{subsec:caseofthespherecenteredatzero} and given by \eqref{eq:phasediscontinuityforsphericalcase}.
\end{remark}

\begin{remark}[Case when $\Gamma$ is off centered]\label{rmk:aieta case is incorrect}\rm
A case considered in \cite[Section 3]{aieta2012reflection} is when a sphere of radius $R$ is centered at a point $(0,0,a)$ with $a> R$, and the authors claim there that it is not possible to find a phase discontinuity on such a sphere so that all rays from the origin are refracted into the vertical direction. 
We believe this claim is in error and in fact, with the method above will show that for each unit $m=(m_1,m_2,m_3)$ with $m_3>0$, there is a phase discontinuity $\psi$ defined in a neighborhood of such a sphere so that its gradient is tangential to the sphere and so that radiation from the origin is refracted into a fixed direction $m$, see Figure \ref{fig:off centered sphere}. 
\begin{figure}
\includegraphics[width=3in]{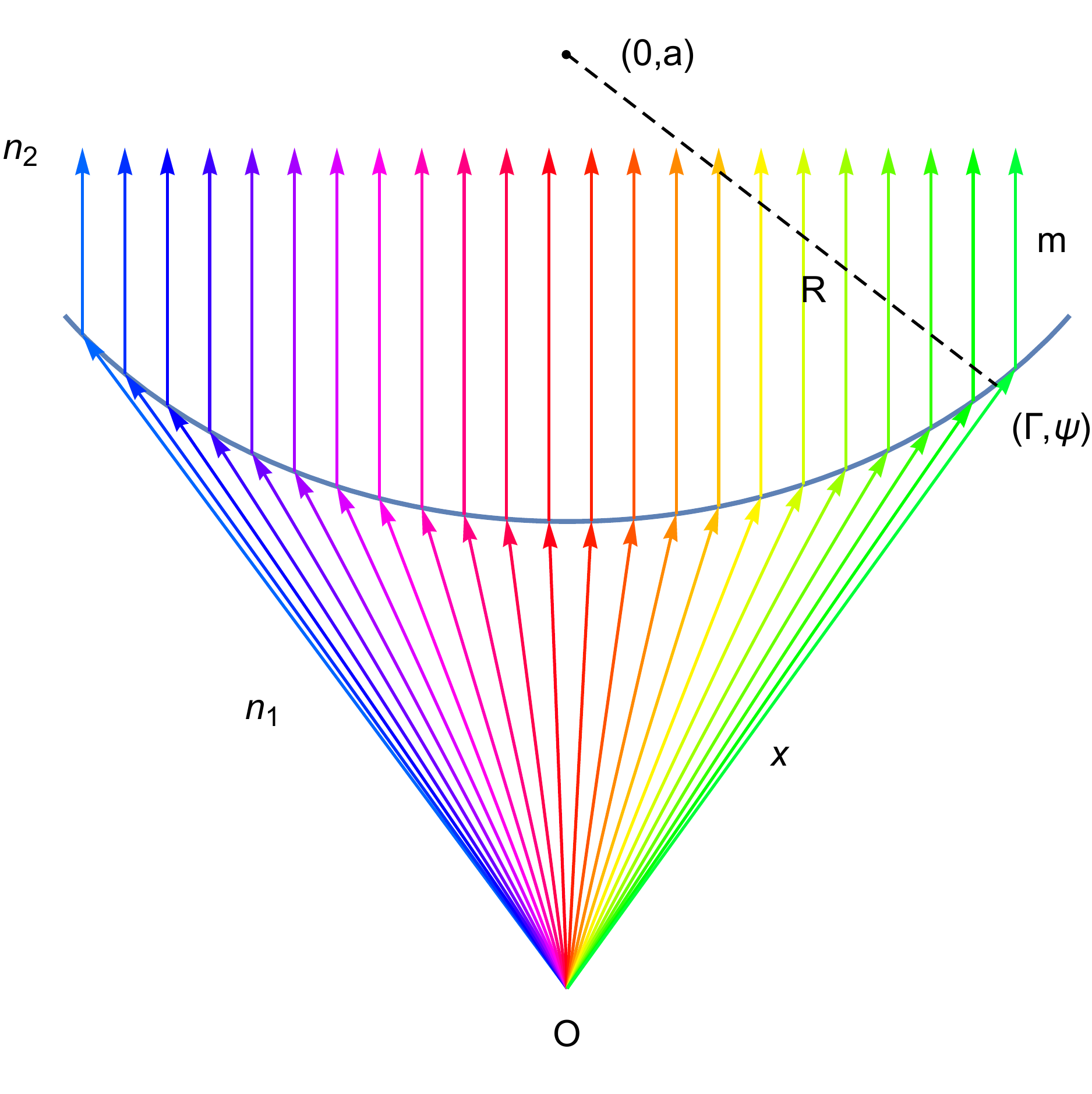}
\caption{Off centered spherical metalens refracting into a fixed direction}
\label{fig:off centered sphere}
\end{figure}
In particular, when $m$ is vertical a phase discontinuity exists.
By reversibility of optical paths, this shows that the conclusion in \cite[Section 3]{aieta2012reflection} is incorrect. 

First, the lower part of the sphere with center at $(0,0,a)$ and radius $R$ is parametrized by the vector
$r(u,v)=\rho(u,v)\,x(u,v)$ with
\[
\rho(u,v)=a\,\cos v-\sqrt{R^2-a^2\,\sin^2 v},
\]
where $0\leq v\leq \arcsin (R/a)$; and the unit normal to the sphere pointing upwards is 
\[
\nu=\dfrac{(0,0,a)-\rho(u,v)\,x(u,v)}{R}.
\]
To show our claim, we need to verify that the determinant in \eqref{formulafor2x2determinantmvar} is not zero.
From \eqref{eq:formulasfordotprodbetweenruandnuvmvar} we obtain by simple calculations that
\begin{align*}
r_{uu}\cdot \nu=-r_u\cdot \nu_u&=\dfrac{1}{R}\left(\sin^2 v \right)\,\rho^2 \\
r_{uv}\cdot \nu=-r_u\cdot \nu_v&=\dfrac{1}{R}\,\rho_u\,\rho_v=0\\
r_{vv}\cdot \nu=-r_v\cdot \nu_v&=\dfrac{1}{R}\,\left( (\rho_v)^2+\rho^2 \right).
\end{align*}
%and so
%\begin{align*}
%\det \left(
%\begin{matrix}
%r_{uu}\cdot \nu & r_{uv}\cdot \nu\\
%r_{uv}\cdot \nu& r_{vv}\cdot \nu
%\end{matrix} \right)
%=
%\dfrac{1}{R^2}\,
%\left(\sin^2 v \right)\,\rho^2\,
%\left((\rho_v)^2+\rho^2 \right).
%\end{align*}
Therefore the determinant in \eqref{formulafor2x2determinantmvar} equals
\begin{align}\label{eq:determinantforspherebis}
\det \left(
\begin{matrix}
F_{uu} & F_{uv}\\
F_{vu}& F_{vv}
\end{matrix} \right)
%&=
%(\sin^2 v)\,\rho^2+
%\dfrac{1}{R}\,B\,(\sin^2 v)\,\rho^3\\
%&\qquad +
%\dfrac{1}{R}\,B\,\rho\,\sin^2 v \left((\rho_v)^2+\rho^2\right)\notag\\
%&\qquad \qquad +
%\dfrac{1}{R^2}\,B^2\,
%\left(\sin^2 v \right)\,\rho^2\,
%\left((\rho_v)^2+\rho^2 \right)\notag\\
&=
\rho\,(\sin^2 v)\, \left(1+\dfrac{B}{R} \,\rho\right)\,
\left(\rho+\dfrac{B}{R}\left( \rho^2+(\rho_v)^2\right) \right),
\end{align}
with \[
B=(x-\kappa\,m)\cdot \nu=\dfrac{1}{R}\,(x-\kappa\,m)\cdot \left((0,0,a)-\rho\,x\right)
=
\dfrac{1}{R}\left(\sqrt{R^2-a^2\,\sin^2 v}-\kappa\,a\,m_3+\kappa\,\rho\,(m\cdot x) \right).
\]
The last determinant is not zero for $u,v$ such that
\begin{equation*}
\sin^2 v\neq 0,\quad 
1+\dfrac{B}{R}\,\rho\neq 0,\quad 
\text{and }\rho+
\dfrac{B}{R}\left( \rho^2+(\rho_v)^2\right)\neq 0.
\end{equation*}
Let us take for example $m=(0,0,1)$, i.e., rays are refracted vertically, then we get
\begin{align*}
B&
%=\dfrac{1}{R}\left(a\,\cos v-\rho-\kappa\,a+\kappa\,\rho\,\cos v \right)\\
%&=
%\dfrac{1}{R}
%\left( 
%\sqrt{R^2-a^2\,\sin^2 v}
%-\kappa\,a
%+
%\kappa\,\cos v \left(a\,\cos v-\sqrt{R^2-a^2\,\sin^2 v} \right)\right)\\
%&=
%\dfrac{1}{R}
%\left( 
%(1-\kappa\,\cos v)\sqrt{R^2-a^2\,\sin^2 v}
%-\kappa\,a
%+
%\kappa\,a\,\cos^2 v \right)\\
%&
=
\dfrac{1}{R}
\left( 
(1-\kappa\,\cos v)\sqrt{R^2-a^2\,\sin^2 v}
-\kappa\,a\,\sin^2 v
\right),
\end{align*}
so $B$ is independent of $u$.
If $v\approx 0$, then $B\approx 1-\kappa$, $\rho\approx a-R$ and $\rho_v\approx 0$, so
\begin{align*}
1+\dfrac{B}{R}\,\rho&\approx 1+(1-\kappa)\left(\dfrac{a}{R}-1 \right)\\
\rho+
\dfrac{B}{R}\left( \rho^2+(\rho_v)^2\right)&\approx (a-R)\left( 1+(1-\kappa)\left(\dfrac{a}{R}-1 \right)\right).
\end{align*}
Recall that $\kappa=n_2/n_1$. If $\kappa<1$, since $a>R$, we obtain that  $1+(1-\kappa)\left(\dfrac{a}{R}-1 \right)\neq 0$.
If $\kappa>1$, then $1+(1-\kappa)\left(\dfrac{a}{R}-1 \right)\neq 0$ if and only if 
$\kappa\neq 1+\dfrac{R}{a-R}$. This shows that in these cases the determinant in
\eqref{eq:determinantforspherebis} is not zero for $v\neq 0$ with $v$ close to zero.
Therefore there exists a phase discontinuity $\psi$, on the sphere centered at $(0,0,a)$ with radius $R$, defined in a neighborhood of each point of the form $\rho(u,v) \, x(u,v)$ with $v$ close to zero.

%This completes the proof that the conclusion in \cite[Section 3]{aieta2012reflection} is incorrect.

\end{remark}

\section{Given a phase discontinuity find an admissible surface}\label{subsec:given the field find the phase discontinuity}
We now turn to the second question proposed at the beginning of Section \ref{sec:Far field uniformly refracting surfaces}, that is, of finding the surface $\Gamma$ when the field $V=(V_1,V_2,V_3)$ is given. 
The unknown surface is given parametrically by 
$$r(u,v)=\rho(u,v)\,x(u,v)
$$ 
where $x(u,v)$ are spherical coordinates as before, and we seek the polar radius $\rho$; the value of $V$ along the surface is $V(r(u,v))$.
From \eqref{eq:generalizedSnell}, $x(u,v)-\kappa\,m- V(r(u,v))$ is a multiple of the normal $\nu$ at $r(u,v)$, so
\[
r_u(u,v)\cdot \left(x(u,v)-\kappa\,m- V(r(u,v)) \right)=0\quad \text{and}\quad
r_v(u,v)\cdot \left(x(u,v)-\kappa\,m- V(r(u,v)) \right)=0.
\] 
We have 
\begin{align*}
r_u(u,v)&=\left[\rho(u,v)\right]_u\,x(u,v)+\rho(u,v)\,x_u(u,v),\\
r_v(u,v)&=\left[\rho(u,v)\right]_v\,x(u,v)+\rho(u,v)\,x_v(u,v),
\end{align*}
so
\begin{align*}
0&=r_u(u,v)\cdot \left(x(u,v)-\kappa\,m- V(r(u,v)) \right)\\
&=\left(\left[\rho(u,v)\right]_u\,x(u,v)+\rho(u,v)\,x_u(u,v)\right)\cdot \left(x(u,v)-\kappa\,m- V(r(u,v)) \right)\\
&=\left[\rho(u,v)\right]_u\left(1-x(u,v)\cdot\left[\kappa\, m+V(r(u,v))\right]\right)
-\rho(u,v)\,x_u(u,v)\cdot \left[\kappa\, m+V(r(u,v))\right],
 \end{align*}
and a similar equation for $r_v$.
That is, $\rho(u,v)$ satisfies the first order nonlinear system of pdes (depending on $V$)\footnote{We are assuming that $1-x(u,v)\cdot\left[\kappa\, m+V\left(\rho(u,v)\,x(u,v)\right)\right]\neq 0$.}:
\begin{equation}\label{eq:odegeneralizedrefraction}
\begin{cases}
\rho_u(u,v) -\dfrac{x_u\cdot \left[\kappa\, m+V\left(\rho(u,v)\,x(u,v)\right)\right]}{1-x(u,v)\cdot\left[\kappa\, m+V\left(\rho(u,v)\,x(u,v)\right)\right]}\,\rho(u,v)=0\\[2ex]
\rho_v(u,v) -\dfrac{x_v\cdot \left[\kappa\, m+V\left(\rho(u,v)\,x(u,v)\right)\right]}{1-x(u,v)\cdot\left[\kappa\, m+V\left(\rho(u,v)\,x(u,v)\right)\right]}\,\rho(u,v)=0.
\end{cases}
\end{equation}
If $F=(F_1,F_2)$ with
\begin{align*}
F_1(u,v,\rho)&=\dfrac{x_u\cdot \left[\kappa\, m+V\left(\rho\,x(u,v)\right)\right]}{1-x(u,v)\cdot\left[\kappa\, m+V\left(\rho\,x(u,v)\right)\right]}\,\rho\\
F_2(u,v,\rho)&=\dfrac{x_v\cdot \left[\kappa\, m+V\left(\rho\,x(u,v)\right)\right]}{1-x(u,v)\cdot\left[\kappa\, m+V\left(\rho\,x(u,v)\right)\right]}\,\rho,
\end{align*}
then \eqref{eq:odegeneralizedrefraction} can be written as
\begin{equation}\label{eq:odegeneralizedrefractionshort}
\nabla \rho=F(u,v,\rho).
\end{equation}
To solve the system \eqref{eq:odegeneralizedrefractionshort} we need an initial condition, say $\rho(u_0,v_0)=\rho_0$, 
and use a result from \cite[Chapter 6, pp. 117-118]{hartman-book-odes}, that is, if 
\begin{equation}\label{eq:Hartmanconditionforsolvability}
\dfrac{\partial F_1}{\partial v}(u,v,\rho) + \dfrac{\partial F_1}{\partial \rho}(u,v,\rho) F_2(u,v,\rho)=
 \dfrac{\partial F_2}{\partial u}(u,v,\rho) +\dfrac{\partial F_2}{\partial \rho}(u,v,\rho)F_1(u,v,\rho)
 \end{equation}
holds for all $(u,v,\rho)$ in an open set $O$, then for each $(u_0,v_0,\rho_0)\in O$ there is neighborhood $U$ of $(u_0,v_0)$ and a unique solution $\rho(u,v)$ defined for $(u,v)\in U$ solving the system \eqref{eq:odegeneralizedrefractionshort} and satisfying $\rho(u_0,v_0)=\rho_0$.

We will see under what circumstances on the field $V$ condition \eqref{eq:Hartmanconditionforsolvability} is satisfied, and therefore the existence of the desired surface $r(u,v)$ will be guaranteed.
Set 
\begin{equation}\label{eq:definitionofW}
W(u,v,\rho)=\kappa\,m+V\left(\rho\,x(u,v)\right),
\end{equation}
then
\[
F_1(u,v,\rho)=\dfrac{x_u\cdot W(u,v,\rho)}{1-x(u,v)\cdot \left[W(u,v,\rho)\right]}\,\rho,
\qquad
F_2(u,v,\rho)=\dfrac{x_v\cdot W(u,v,\rho)}{1-x(u,v)\cdot \left[W(u,v,\rho)\right]}\,\rho.
\]
We have
%{\small
\begin{align*}
\dfrac{\partial F_1}{\partial v}
&=
\left\{\left(x_{uv}\cdot W+x_u\cdot W_v \right)(1-x\cdot W)^{-1}+\left(x_v\cdot W+x\cdot W_v\right)
(x_u\cdot W)\,(1-x\cdot W)^{-2}\right\}\,\rho\\
\dfrac{\partial F_2}{\partial u}
&=
\left\{\left(x_{vu}\cdot W+x_v\cdot W_u \right)(1-x\cdot W)^{-1}+\left(x_u\cdot W+x\cdot W_u\right)
(x_v\cdot W)\,(1-x\cdot W)^{-2}\right\}\,\rho\\
\dfrac{\partial F_1}{\partial \rho}
&=
(x_u\cdot W)\,(1-x\cdot W)^{-1}+
\left\{\left(x_u\cdot W_\rho \right)(1-x\cdot W)^{-1}
+
(x_u\cdot W)\,(x\cdot W_\rho)\,(1-x\cdot W)^{-2}\right\}\,\rho\\
\dfrac{\partial F_2}{\partial \rho}
&=
(x_v\cdot W)\,(1-x\cdot W)^{-1}+
\left\{\left(x_v\cdot W_\rho \right)(1-x\cdot W)^{-1}
+
(x_v\cdot W)\,(x\cdot W_\rho)\,(1-x\cdot W)^{-2}\right\}\,\rho.
\end{align*}
%}
Hence
\begin{dmath*}
\dfrac{\partial F_1}{\partial v}-\dfrac{\partial F_2}{\partial u} =
\left\{\left(x_u\cdot W_v -x_v\cdot W_u\right)(1-x\cdot W)^{-1}
+
\left((x\cdot W_v)\,(x_u\cdot W)-(x\cdot W_u)\,(x_v\cdot W) \right) \,(1-x\cdot W)^{-2}\right\}\,\rho
\end{dmath*}
%}
%%{\small
%\begin{align*}
%\dfrac{\partial F_1}{\partial v}-\dfrac{\partial F_2}{\partial u}
%&= \\
%&=\left\{\left(x_u\cdot W_v -x_v\cdot W_u\right)(1-x\cdot W)^{-1}
%+
%\left((x\cdot W_v)\,(x_u\cdot W)-(x\cdot W_u)\,(x_v\cdot W) \right) \,(1-x\cdot W)^{-2}\right\}\,\rho
%\end{align*}
%}
and
\begin{dmath*}
\dfrac{\partial F_1}{\partial \rho}F_2-
 \dfrac{\partial F_2}{\partial \rho}F_1\\
 =
 \left[(x_u\cdot W)\,(1-x\cdot W)^{-1}+
\left\{\left(x_u\cdot W_\rho \right)(1-x\cdot W)^{-1}
+
(x_u\cdot W)\,(x\cdot W_\rho)\,(1-x\cdot W)^{-2}\right\}\,\rho \right]\,
(x_v\cdot W)\,(1-x\cdot W)^{-1}\\
-
\left[(x_v\cdot W)\,(1-x\cdot W)^{-1}+
\left\{\left(x_v\cdot W_\rho \right)(1-x\cdot W)^{-1}
+
(x_v\cdot W)\,(x\cdot W_\rho)\,(1-x\cdot W)^{-2}\right\}\,\rho \right](x_u\cdot W)\,(1-x\cdot W)^{-1} \\
=
\left((x_u\cdot W_\rho)\,(x_v\cdot W)- (x_v\cdot W_\rho)\,(x_u\cdot W)\right)\,(1-x\cdot W)^{-2}\,\rho.\end{dmath*}
%}

%{\tiny 
%\begin{align*}
%&\dfrac{\partial F_1}{\partial \rho}F_2-
% \dfrac{\partial F_2}{\partial \rho}F_1\\
% &=
% \left[(x_u\cdot W)\,(1-x\cdot W)^{-1}+
%\left\{\left(x_u\cdot W_\rho \right)(1-x\cdot W)^{-1}
%+
%(x_u\cdot W)\,(x\cdot W_\rho)\,(1-x\cdot W)^{-2}\right\}\,\rho \right]\,
%(x_v\cdot W)\,(1-x\cdot W)^{-1}\\
%&-
%\left[(x_v\cdot W)\,(1-x\cdot W)^{-1}+
%\left\{\left(x_v\cdot W_\rho \right)(1-x\cdot W)^{-1}
%+
%(x_v\cdot W)\,(x\cdot W_\rho)\,(1-x\cdot W)^{-2}\right\}\,\rho \right](x_u\cdot W)\,(1-x\cdot W)^{-1} \\
%&=
%\left((x_u\cdot W_\rho)\,(x_v\cdot W)- (x_v\cdot W_\rho)\,(x_u\cdot W)\right)\,(1-x\cdot W)^{-2}\,\rho.\end{align*}
%}
Therefore \eqref{eq:Hartmanconditionforsolvability} holds if
\begin{align*}
&\dfrac{\partial F_1}{\partial v}-\dfrac{\partial F_2}{\partial u}
+\dfrac{\partial F_1}{\partial \rho}F_2-
 \dfrac{\partial F_2}{\partial \rho}F_1\\
 &=
 \left\{\left(x_u\cdot W_v -x_v\cdot W_u\right)(1-x\cdot W)^{-1}
+
\left((x\cdot W_v)\,(x_u\cdot W)-(x\cdot W_u)\,(x_v\cdot W) \right) \,(1-x\cdot W)^{-2}\right\}\,\rho\\
&\qquad +
\left((x_u\cdot W_\rho)\,(x_v\cdot W)- (x_v\cdot W_\rho)\,(x_u\cdot W)\right)\,(1-x\cdot W)^{-2}\,\rho=0.
  \end{align*}
Since we assume $1-x\cdot W\neq 0$ and $\rho>0$, this is equivalent to
\begin{align*}
&\left(x_u\cdot W_v -x_v\cdot W_u\right)(1-x\cdot W)
+
\left((x\cdot W_v)\,(x_u\cdot W)-(x\cdot W_u)\,(x_v\cdot W) \right)\\
&\qquad +
\left((x_u\cdot W_\rho)\,(x_v\cdot W)- (x_v\cdot W_\rho)\,(x_u\cdot W)\right)=0,
\end{align*}
that is,
%{\small
\begin{dmath}\label{eq:conditionequivalenttoHartman}
\left(x_u\cdot W_v -x_v\cdot W_u\right)(1-x\cdot W)
+
\left(\left((x\cdot W_v)-(x_v\cdot W_\rho)\right)\,(x_u\cdot W)-\left( (x\cdot W_u)-(x_u\cdot W_\rho)\right)\,(x_v\cdot W) \right)=0.
\end{dmath}
%}
We have 
\begin{align*}
W_u&=\rho\,\left(\nabla V_1\cdot x_u,\nabla V_2\cdot x_u,\nabla V_3\cdot x_u \right)\\
W_v&=\rho\,\left(\nabla V_1\cdot x_v,\nabla V_2\cdot x_v,\nabla V_3\cdot x_v \right)\\
W_\rho&=\left(\nabla V_1\cdot x,\nabla V_2\cdot x,\nabla V_3\cdot x \right).
\end{align*}
Now
\[
x\cdot W_v=\rho\,\sum_{k=1}^3 x_k\,(\nabla V_k\cdot x_v)
=
\rho\,\sum_{k=1}^3 x_k\,\sum_{j=1}^3 \dfrac{\partial V_k}{\partial y_j} (x_j)_v
=
\rho\,\sum_{k,j=1}^3  \dfrac{\partial V_k}{\partial y_j} (x_j)_v\,x_k.
\]
If we let
\[
A=\left(
\begin{matrix}
\dfrac{\partial V_1}{\partial y_1} & \dfrac{\partial V_1}{\partial y_2} & \dfrac{\partial V_1}{\partial y_3} \\[2ex]
\dfrac{\partial V_2}{\partial y_1} & \dfrac{\partial V_2}{\partial y_2} & \dfrac{\partial V_2}{\partial y_3} \\[2ex]
\dfrac{\partial V_3}{\partial y_1} & \dfrac{\partial V_3}{\partial y_2} & \dfrac{\partial V_3}{\partial y_3} \\
\end{matrix}
\right),
\]
then
\[
x\cdot W_v=\rho\,x\,A\,(x_v)^t
\]
where $x,x_v$ are row vectors and $t$ denotes the transpose.
Similarly
\begin{align*}
x\cdot W_u&=\rho\,x\,A\,(x_u)^t\quad
x_u\cdot W_v=\rho\,x_u\,A\,(x_v)^t\quad\\
x_v\cdot W_u&=\rho\,x_v\,A\,(x_u)^t\quad
x_u\cdot W_\rho=x_u\,A\,(x)^t\quad
x_v\cdot W_\rho=x_v\,A\,(x)^t.
\end{align*}
Suppose $V=\nabla \psi$, then $A=\nabla^2\psi$ is a symmetric matrix, so
\[
x_u\cdot W_v=x_v\cdot W_u
\]
\[
(x\cdot W_v)-(x_v\cdot W_\rho)=(\rho-1)\,x\,A\,(x_v)^t=\dfrac{\rho-1}{\rho}\,(x\cdot W_v)
\]
\[
(x\cdot W_u)-(x_u\cdot W_\rho)=(\rho-1)\,x\,A\,(x_u)^t=\dfrac{\rho-1}{\rho}\,(x\cdot W_u)\]
and \eqref{eq:conditionequivalenttoHartman} reads
\begin{equation}\label{eq:conditionequivalenttoHartmanbis}
(\rho-1)\,
\left\{(x\,A\,(x_v)^t)\, (x_u\cdot W)-(x\,A\,(x_u)^t)\, (x_v\cdot W)\right\}=0,\footnote{Since $(x\cdot W)_u=x_u\cdot W+\rho\,x\,A\,(x_u)^t$ and similarly for $(x\cdot W)_v$, this condition can be re-written as $(\rho-1)\left\{x\,A\,(x_v)^t\, (x\cdot W)_u-x\,A\,(x_u)^t\, 
(x\cdot W)_v\right\}=0$.}
\end{equation}
which can be written as
\[
\det
\left( 
\begin{matrix}
x\,A\,(x_u)^t & x\,A\,(x_v)^t\\
x_u\cdot W & x_v\cdot W
\end{matrix}
\right)
=
\det
\left( 
\begin{matrix}
x_u\cdot Ax & x_v\cdot Ax\\
x_u\cdot W & x_v\cdot W
\end{matrix}
\right)=
0.
\]
From the Cauchy-Binet formula for cross products\footnote{$(a\times b)\cdot (c\times d)=(a\cdot c)(b\cdot d)-(a\cdot d)(b\cdot c)$.}, this means that 
\[(x_u\times x_v)\cdot (Ax\times W)=0
\]
and since $x_u\times x_v\parallel x$, \eqref{eq:conditionequivalenttoHartmanbis} is equivalent to the following geometric condition: 
\begin{equation}\label{eq:Hartmangeometriccondition}
x\cdot (Ax\times W)=0.\footnote{Equivalently 
$W\cdot (Ax\times x)=Ax\cdot (x\times W)=0$.}
\end{equation}

{\it Therefore, if the field $V=\nabla \psi$, $W$ is given in \eqref{eq:definitionofW}, and \eqref{eq:conditionequivalenttoHartmanbis} (or equivalently \eqref{eq:Hartmangeometriccondition}) holds in an open set $O$ in the variables $(\rho,u,v)$, then for each $(\rho_0,u_0,v_0)\in O$ the system \eqref{eq:odegeneralizedrefraction} has a unique solution $\rho(u,v)$ defined in a neighborhood of $(u_0,v_0)$ and satisfying the initial condition $\rho(u_0,v_0)=\rho_0$.}
Notice that if $V=V_0$ is a constant field, then $A=0$ and so \eqref{eq:conditionequivalenttoHartmanbis} obviously holds.
In this case, \eqref{eq:odegeneralizedrefraction} can be easily integrated and the solution is 
\[
\rho(u,v)=\dfrac{C_1}{1-x(u,v)\cdot (\kappa\,m+V_0)}+C_2
\] 
with $C_i$ constants.

Notice also that with the choice $V$ as in \eqref{eq:fieldforplanex=a}, with $h\neq 0$ so $1-x\cdot W\neq 0$, 
the system of equations \eqref{eq:odegeneralizedrefraction} becomes
\[
\begin{cases}
\rho_u(u,v) -\dfrac{\sin u}{\cos u}\,\rho(u,v)=0\\
\\
\rho_v(u,v) +\dfrac{\cos v}{\sin v}\,\rho(u,v)=0,
\end{cases}
\]
whose solution is $\rho(u,v)=\dfrac{C}{\cos u\,\sin v}$, where the constant $C$ is determined by the point where the solution passes through. This is in agreement with 
\eqref{eq:parametricequationofplane}.

\section{Near field refracting metasurfaces}\label{sec:near field case}
\setcounter{equation}{0}
%The starting point is again \eqref{eq:generalizedSnell}. 
The near field case can be regarded as a special case from Section \ref{sec:m variable}
when the vector field $m(u,v)$ points towards a fixed point $Q$, and therefore the method from that section can be used to derive conditions for the existence of the desired metasurface.
In fact, if the surface $\Gamma$ is parametrized by $r(u,v)$ and $m(u,v)=\dfrac{Q-r(u,v)}{|Q-r(u,v)|}$, then it is easy to see that the compatibility condition \eqref{eq:compatibility condition between m and Gamma} holds.
The existence of the phase discontinuity then follows when the determinant in \eqref{formulafor2x2determinantmvar} is not zero.

\begin{figure}[htp]
\begin{center}
    \subfigure[$ $]{\label{fig:near field case of the plane}\includegraphics[width=2.9in]{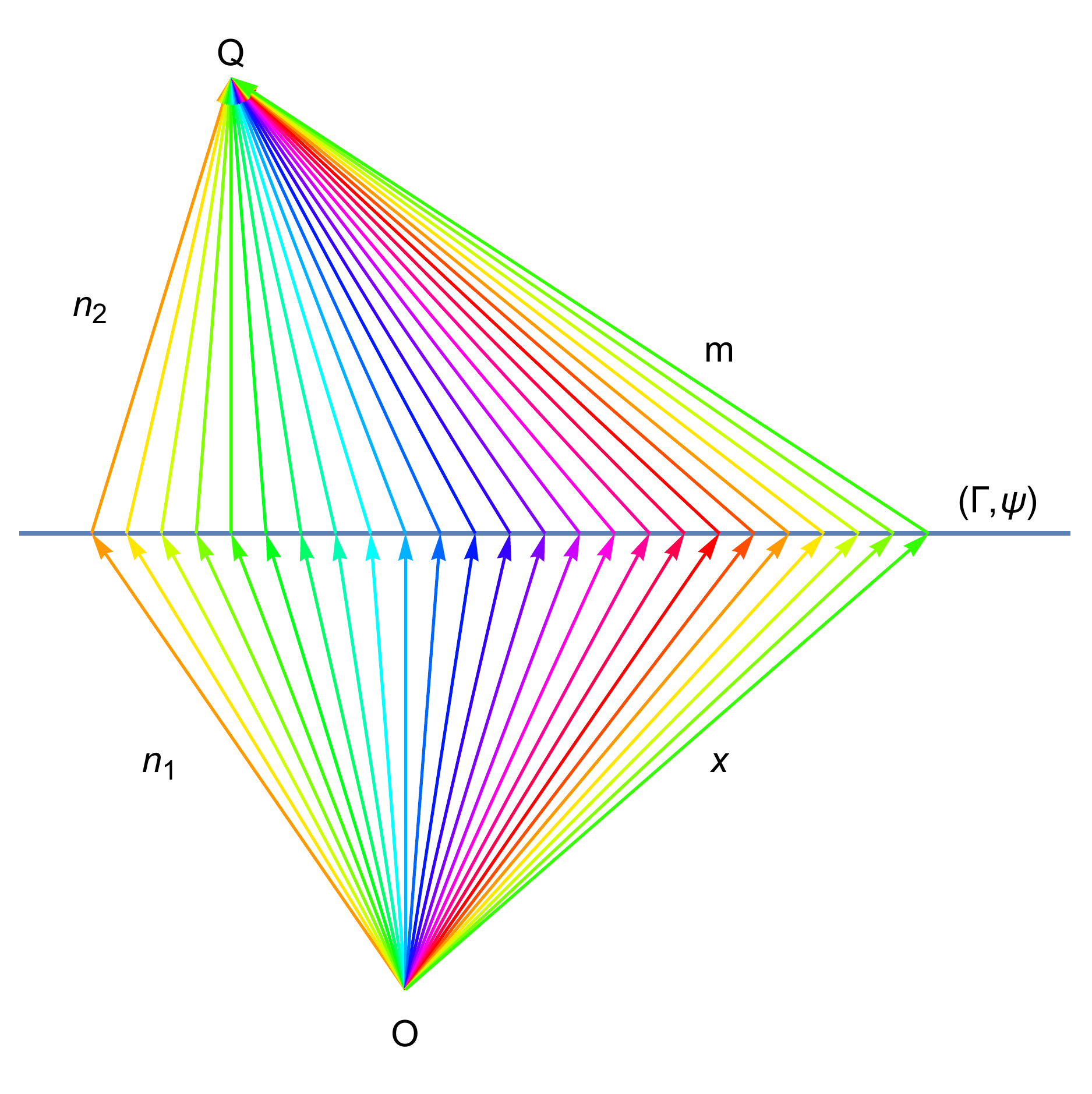}}
    \subfigure[$ $]{\label{fig:near field case of the sphere}\includegraphics[width=2.9in]{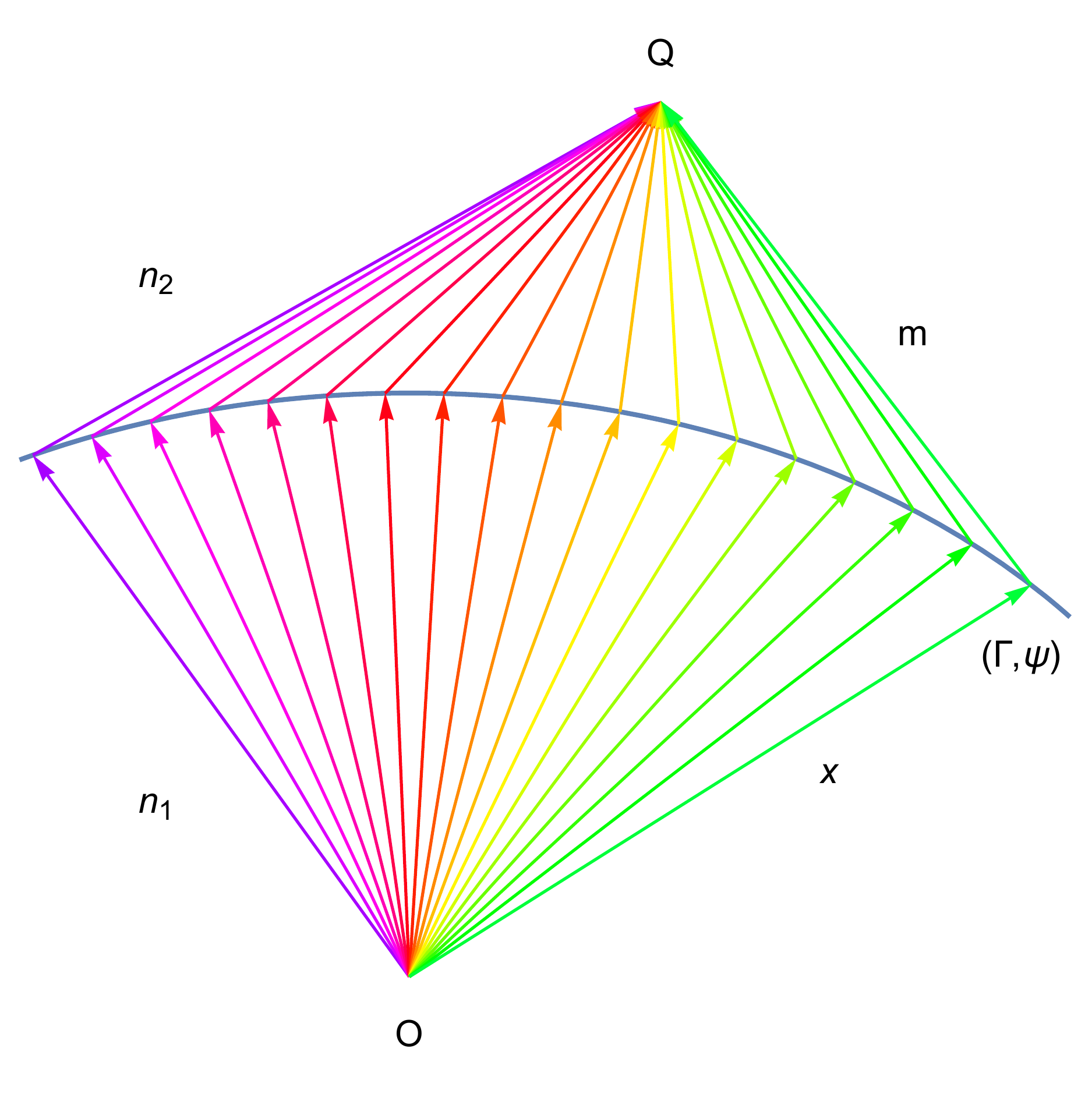}} 
%    \subfigure[After Sobel edge detection]{\label{fig:edge-c}\includegraphics[scale=1]{SeveralCartesianOvalskappa<1.pdf}}
\end{center}
  \caption{Planar and spherical metalenses in the near field}
  \label{fig:cases of plane and sphere near field}
\end{figure}

However, the phase discontinuities in the planar and spherical cases can be obtained explicitly as follows; see Figure \ref{fig:cases of plane and sphere near field}.
\subsection{Case of a plane interface}\label{subsec:near field plane case}
Let $O$ be the origin in medium $I$ with index $n_1$ and let $Q=(q_1,q_2,q_3)$ be a point in medium $II$ with index $n_2$. Denote by $\Gamma$ the plane with equation $x_1=a$ so that it separates the points $O$ and $Q$.
We find the field $V$ so that rays from $O$ are refracted into $Q$.
We know from Section \ref{subsec:case of the plane} that $\Gamma$ is given parametrically by \eqref{eq:parametricequationofplane}; the normal $\nu=(1,0,0)$.
So we seek $V$ such that \eqref{eq:generalizedSnell} holds. Since the refracted vector from each point $r(u,v)$ on the plane interface to the point $Q$ has unit direction $\dfrac{Q-r(u,v)}{|Q-r(u,v)|}$, $V$ must satisfy
\begin{align*}
\cos u\,\sin v-\kappa\,\dfrac{q_1-a}{|Q-r(u,v)|}&=\lambda + V_1\\
\sin u\,\sin v-\kappa\,\dfrac{q_2-a\,\tan u}{|Q-r(u,v)|}&= V_2\\
\cos v-\kappa\,\dfrac{q_3-a/\cos u \tan v}{|Q-r(u,v)|}&= V_3.
\end{align*}
Re writing these equations in rectangular coordinates yields
\begin{align*}
\dfrac{a}{\sqrt{a^2+x_2^2+x_3^2}}-\kappa\,\left.\dfrac{q_1-a}{|Q-(x_1,x_2,x_3)|}\right|_{x_1=a}&=\lambda + V_1\\
\dfrac{x_2}{\sqrt{a^2+x_2^2+x_3^2}}-\kappa\,\left.\dfrac{q_2-x_2}{|Q-(x_1,x_2,x_3)|}\right|_{x_1=a}&= V_2\\
\dfrac{x_3}{\sqrt{a^2+x_2^2+x_3^2}}-\kappa\,\left.\dfrac{q_3-x_3}{|Q-(x_1,x_2,x_3)|}\right|_{x_1=a}&= V_3.
\end{align*}
Therefore, $V_i$, $i=1,2,3$, are determined:
\begin{align*}
V_1(a,x_2,x_3)&=\left. \partial_{x_1}\left(\sqrt{x_1^2+x_2^2+x_3^2} \right)\right|_{x_1=a}+\kappa\,\left.\dfrac{\partial}{\partial x_1}|Q-(x_1,x_2,x_3)|\right|_{x_1=a}-\lambda\\
V_2(a,x_2,x_3)&=\left. \partial_{x_2}\left(\sqrt{x_1^2+x_2^2+x_3^2} \right)\right|_{x_1=a}+\kappa\,\left.\dfrac{\partial}{\partial x_2}|Q-(x_1,x_2,x_3)|\right|_{x_1=a}\\
V_3(a,x_2,x_3)&=\left. \partial_{x_3}\left(\sqrt{x_1^2+x_2^2+x_3^2} \right)\right|_{x_1=a}+\kappa\,\left.\dfrac{\partial}{\partial x_3}|Q-(x_1,x_2,x_3)|\right|_{x_1=a},
\end{align*}
where $\lambda$ is chosen arbitrarily.
Notice that if we let 
\[
\psi(x_1,x_2,x_3)=\sqrt{x_1^2+x_2^2+x_3^2} +\kappa\,|Q-(x_1,x_2,x_3)|
\]
and choose $\lambda=0$, then $V=\nabla \psi$, and so the plane with the phase discontinuity function $\psi$ does the desired refraction job.

\subsection{Case of a spherical interface}\label{subsec:near field spherical case}
If $\Gamma$ is the sphere of radius $R$ centered at the origin, that is, $r(u,v)=R\,x(u,v)$, then the normal $\nu=x$, and from \eqref{eq:generalizedSnell} we get
\[
\left(x-\kappa\,\dfrac{Q-r(u,v)}{|Q-r(u,v)|}-V\right)\times x=0.
\]
As before taking cross product with $x$ yields
\[
V+\kappa\,\dfrac{Q-r(u,v)}{|Q-r(u,v)|}-\left( \kappa \,\left(\dfrac{Q-r(u,v)}{|Q-r(u,v)|}\cdot x\right)+V\cdot x\right)\,x=0.
\]
Assuming $V$ is tangential to the sphere,
\[
V=
-\kappa\,\dfrac{Q-r(u,v)}{|Q-r(u,v)|}+ \kappa \,\left(\dfrac{Q-r(u,v)}{|Q-r(u,v)|}\cdot x\right)\,x.
\]
%hence
%\[
%V(R\,x)
%=
%\kappa\,\left.D_z\left( |Q-z|\right)\right|_{z=Rx}- \kappa \,\left( \left.D_z\left( |Q-z|\right)\right|_{z=Rx}\cdot x\right)
%\,x,\text{ with $|x|=1$}.
%\]
If $V(Rx(u,v))=(\nabla \psi)(Rx(u,v))$, then
\begin{equation}\label{eq:equationforgradpsionthespherenearfield}
\psi_{x_j}(Rx(u,v))=-\kappa\,\dfrac{q_j-Rx_j(u,v)}{|Q-Rx(u,v)|}+\kappa\left(\dfrac{Q-Rx(u,v)}{|Q-Rx(u,v)|}\cdot x\right) \,x_j,\qquad j=1,2,3.
\end{equation}
Hence  
\begin{align}\label{eq:derivative in u}
\dfrac{\partial }{\partial u}\left( \psi(Rx(u,v))\right)
&=
(\nabla \psi)(Rx(u,v))\cdot Rx_u
=
-\kappa\,R\,\dfrac{Q-Rx(u,v)}{|Q-Rx(u,v)|}\cdot x_u,
%=-\kappa\,R\,\dfrac{Q\cdot x_u}{|Q-Rx(u,v)|},
%&=\left( -R\kappa\,\dfrac{Q-Rx(u,v)}{|Q-Rx(u,v)|}\cdot x\right)_u + \kappa\,R\left(\dfrac{Q-Rx(u,v)}{|Q-Rx(u,v)|}\right)_u\cdot x,
\end{align}
and similarly 
\begin{align}\label{eq:derivative in v}
\dfrac{\partial }{\partial v}\left( \psi (Rx(u,v))\right)
&=
-\kappa\,R\,\dfrac{Q-Rx(u,v)}{|Q-Rx(u,v)|}\cdot x_v,
%=
%-\kappa\,R\,\dfrac{Q\cdot x_v}{|Q-Rx(u,v)|},
\end{align}
since $x\cdot x_u=x\cdot x_v=0$.
Since $\psi$ is assumed $C^2$, we get
\begin{equation}\label{eq:mix derivatives}
\left(\dfrac{Q-Rx(u,v)}{|Q-Rx(u,v)|}\right)_u\cdot x_v=\left(\dfrac{Q-Rx(u,v)}{|Q-Rx(u,v)|}\right)_v\cdot x_u.
\end{equation}
Integrating \eqref{eq:derivative in u} in $u$ yields
\[
\psi(Rx(u,v))
=
-\kappa\,R\,\int \dfrac{Q-Rx(u',v)}{|Q-Rx(u',v)|}\cdot x_u(u',v)\,du'+h(v),
\]
for some function $h$. To calculate $h$, we differentiate the integral with respect to $v$ and use \eqref{eq:mix derivatives}: 
\begin{align*}
&\dfrac{\partial }{\partial v}\left( \psi (Rx(u,v))\right)\\
&=
-\kappa\,R\,\int \dfrac{\partial }{\partial v}\left(\dfrac{Q-Rx(u',v)}{|Q-Rx(u',v)|}\cdot x_u(u',v)\right)\,du'+h'(v)\\
&=
-\kappa\,R\,\int \left\{\dfrac{\partial }{\partial v}\left(\dfrac{Q-Rx(u',v)}{|Q-Rx(u',v)|}\right)\cdot x_u(u',v)+\dfrac{Q-Rx(u',v)}{|Q-Rx(u',v)|}\cdot x_{uv}(u',v) \right\}\,du'+h'(v)\\
&=
-\kappa\,R\,\int \left\{\dfrac{\partial }{\partial u}\left(\dfrac{Q-Rx(u',v)}{|Q-Rx(u',v)|}\right)\cdot x_v(u',v)+\dfrac{Q-Rx(u',v)}{|Q-Rx(u',v)|}\cdot x_{vu}(u',v) \right\}\,du'+h'(v)\\
&=
-\kappa\,R\,\int \dfrac{\partial }{\partial u}\left(\dfrac{Q-Rx(u',v)}{|Q-Rx(u',v)|}\cdot x_v(u',v)\right)\,du'+h'(v)\\
&=
-\kappa\,R\,\left(\dfrac{Q-Rx(u,v)}{|Q-Rx(u,v)|}\cdot x_v(u,v)\right)+h'(v)
\end{align*}
which implies $h'(v)=0$ from \eqref{eq:derivative in v}.
Therefore, the phase discontinuity $\psi$ on the sphere satisfies
\begin{align*}
\psi(Rx(u,v))
&=
-\kappa\,R\,\int \dfrac{Q-Rx(u',v)}{|Q-Rx(u',v)|}\cdot x_u(u',v)\,du'+C \\
&= \kappa\,\int \partial_{u}(|Q-Rx(u',v)|)\,du'+C=  \kappa\,|Q-Rx(u,v)|+C
\end{align*}
with $C$ a constant. Writing this in rectangular coordinates yields 
\[
\psi(R(z_1,z_2,z_3))
= \kappa\,|Q-R(z_1,z_2,z_3)|+C ,\qquad \text{ for $|(z_1,z_2,z_3)|=1$.}
\]
We now define $\psi$ on a neighborhood of $|z|=R$ so that \eqref{eq:equationforgradpsionthespherenearfield} holds. Let 
\begin{equation}\label{eq:phase discontinuity near field sphere}
\psi(z)=\kappa\,\left|Q-R\frac{z}{|z|}\right|+C,\,\text{ for $R-\epsilon<|z|<R+\epsilon$}.
\end{equation}
We have 
\[
\nabla \psi(z)
=
-\kappa R \frac{Q-R\frac{z}{|z|}}{\left|Q-R\frac{z}{|z|}\right|}\,\dfrac{1}{|z|}+\kappa R\left(\frac{Q-R\frac{z}{|z|}}{\left|Q-R\frac{z}{|z|}\right|}\cdot \frac{z}{|z|} \right)\frac{z}{|z|^2},
\]
so for $z=Rx$, with $|x|=1$, we obtain
\[
\nabla \psi(Rx)
=
-\kappa \, \frac{Q-Rx}{|Q-Rx|}+\kappa \left(\frac{Q-Rx}{|Q-Rx|}\cdot x \right)x
\]
as desired. Therefore the phase discontinuity $\psi$ in \eqref{eq:phase discontinuity near field sphere} has gradient tangential to the sphere and can be placed on the spherical interface $|z|=R$ so that all rays from the origin are refracted into the point $Q$.

\section{Conclusion}
A rigorous mathematical foundation of general metasurfaces is provided. The 
starting point is the derivation of a generalized Snell's law in the presence of a phase discontinuity using wavefronts. This is used also to derive all possible critical angles.
We solve, under appropriate curvature type conditions on the surface $\Gamma$, the problem of finding a phase discontinuity, so that the pair (surface and phase discontinuity) refracts light in a desired manner. When a phase discontinuity is given, we derive conditions so that a surface
is admissible for that phase discontinuity in the far field setting. 
Extensions to the case when the far field is a set of variable directions are given, and examples and explicit calculations of phase discontinuities are also provided.
%of illustration
%We have explicitly calculated the phase discontinuity when $\Gamma$ is a plane or a sphere in the far field case.
The near field case is also studied.
% when $\Gamma$ is a plane or a sphere.

%\section{QUESTIONS}
%\begin{itemize}
%\item equation in dimension 3
%
%\item What $V$'s are relevant in the applications? WHAT KIND OF $\psi$'S ARE USED IN THE APPLICATIONS? THERE ARE CASES OF $\psi$ FOR WHICH ONE CANNOT HAVE GENERALIZED REFRACTION, SEE \eqref{eq:conditionondiscriminant}.
%WHAT ARE THE MATHEMATICAL LIMITATIONS OF THESE CONSTRUCTIONS?
%
%\item Similarly find the surface (or curve) for the near field, we should get a generalized oval.
%
%\item What are the physical conditions to obtain refraction, i.e., what is the role of internal reflection? \textcolor{blue}{This could be interesting because somehow (????) there are two critical angles for total internal reflection; see \cite[Equation (3)]{yu2011light}. For reflection the critical angle is found in \cite[Equation (5)]{yu2011light}. } WHY ARE THERE TWO CRITICAL ANGLES?
%
%\item Prove rigorously the generalized Snell law using wave fronts. \textcolor{blue}{See Section 3.}
%
%\item \textcolor{blue}{What is the refracted direction $m$ in the case of reflection $n_1 = n_2$? Either both positive or both negative? }
%
%
%\end{itemize}

\newcommand{\etalchar}[1]{$^{#1}$}

%\bibliography{myBib}
%%\bibliography{monamp}
%\bibliographystyle{alpha}

\end{document}